\documentclass[11pt]{article}
\usepackage{geometry}
\usepackage{amsmath}
\usepackage{hyperref}
\usepackage{amssymb}
\usepackage{mathtools}
\usepackage{subcaption}
\usepackage{graphicx}
\usepackage{natbib}
\usepackage{multicol}
\usepackage{algorithm,algpseudocode}

\usepackage{authblk}
\newtheorem{remark}{Remark}%

\usepackage{bm}

\newcommand{\bX}{\bm{X}}

\newcommand{\bZ}{\bm{Z}}

\newcommand{\bU}{\bm{U}}
\newcommand{\bS}{\bm{S}}
\newcommand{\bI}{\bm{I}}

\newcommand{\bb}{\bm{b}}

\newcommand{\bP}{\bm{P}}
\newcommand{\bJ}{\bm{J}}
\newcommand{\bK}{\bm{K}}
\newcommand{\bA}{\bm{A}}
\newcommand{\bW}{\bm{W}}

\newcommand{\ba}{\bm{a}}

\newcommand{\bF}{\bm{F}}

\newcommand{\bsig}{\bm{\Sigma}}
\newcommand{\bu}{\bm{u}}

\newcommand{\beps}{\bm{\epsilon}}
\newcommand{\bsigma}{\bm{\sigma}}

\newcommand{\E}{\mathbb{E}}
\newcommand{\blam}{\bm{\Lambda}}
\newcommand{\btheta}{\bm{\theta}}
\newcommand{\bomega}{\bm{\Omega}}

\newcommand{\bL}{\bm{L}}

\DeclareMathOperator*{\argmin}{arg\,min}

\date{}
\begin{document}

\title{The Sparse Dynamic Factor Model: A Regularised Quasi-Maximum Likelihood Approach}


\author{Luke Mosley}

\author{Tak-Shing T. Chan}

\author{Alex Gibberd\footnote{(a.gibberd@lancaster.ac.uk)}}

\affil{Department of Mathematics and Statistics, Lancaster University, Lancaster, LA1 4YW, United Kingdom}

\maketitle

\abstract{The concepts of sparsity, and regularised estimation, have proven useful in many high-dimensional statistical applications. Dynamic factor models (DFMs) provide a parsimonious approach to modelling high-dimensional time series, however, it is often hard to interpret the meaning of the latent factors. This paper formally introduces a class of sparse DFMs whereby the loading matrices are constrained to have few non-zero entries, thus increasing interpretability of factors. We present a regularised M-estimator for the model parameters, and construct an efficient expectation maximisation algorithm to enable estimation. Synthetic experiments demonstrate consistency in terms of estimating the loading structure, and superior predictive performance where a low-rank factor structure may be appropriate. The utility of the method is further illustrated in an application forecasting electricity consumption across a large set of smart meters.}

\section{Introduction}\label{sec:1}


Originally formalised by \cite{geweke1977dynamic}, the premise of the Dynamic Factor Model (DFM) is to assume that the common dynamics of a large number of stationary zero-mean time series $\bX_t = (X_{1,t}, \dots, X_{p,t})^\top$ stem from a relatively small number of unobserved (latent) factors $\bF_t = (F_{1,t},\dots,F_{r,t})^\top$ where $r \ll p$ through the linear system 
\begin{equation}
    \label{eq:measurement}
    \bX_t = \blam \bF_t + \beps_t \, ,
\end{equation}
for observations $t=1,\dots,n$. The matrix $\blam$ provides a direct link between each factor in $\bF_t$ and each variable in $\bX_t$. The larger the loading 
$ \lvert\Lambda_{i,j}\rvert $ for variable $i$ and factor $j$, the more correlated this variable is with the factor. The common component $\bm{\chi}_t = \blam\bF_t$ captures the variability in the time-series variables that is due to the common factors, while the idiosyncratic errors $\beps_t = (\epsilon_{1,t},\dots,\epsilon_{p,t})^\top$ capture the features that are specific to individual series, such as measurement error. What makes the factor model in (\ref{eq:measurement}) a \emph{dynamic} factor model is the assumption that the factors, and possibly the idiosyncratic errors may be temporally dependent, i.e., are time series themselves.


Arguably, it was the application of \cite{sargent1977business} showing how just two dynamic factors were able to explain the majority of variance in headline US macroeconomic variables that initiated the DFMs popularity. The DFM is nowadays ubiquitous within the economic statistics community, with applications in nowcasting/forecasting \citep{giannone2008nowcasting,banbura2010nowcasting,foroni2014comparison}, constructing economic indicators \citep{mariano2010coincident, grassi2015euromind}, and counterfactual analysis \citep{harvey1996intervention, luciani2015monetary}. Examples in other domains include psychology \citep{molenaar1985dynamic, fisher2015toward}, the energy sector \citep{wu2013new, lee2016load} and many more, see \cite{stock2011dynamic} and \cite{poncela2021factor} for detailed surveys of the literature. 


The DFM can be used in both an exploratory (inferential) setting, as well as a predictive (forecasting) mode. When dealing with the former its use is analogous to how one might apply Principal Component Analysis (PCA) to understand the directions of maximum variation in a dataset, of course, the DFM does not just describe the cross-correlation structure, like PCA, but also the autocovariance. The loadings matrix $\blam$ is usually used to assess how one should interpret a given (estimated) factor. Unfortunately, as in PCA, the interpretation of the factors in a traditional DFM is blurred as all variables are loaded onto all factors. 

\subsection*{Our Contributions}

This paper seeks to bring modern tools from sparse modelling and regularised estimation to bear on the DFM. Specifically, we formalise a class of sparse factor models whereby only a subset of the factors will be active for a given variable---we assume the matrix $\blam$ is sparse. Unlike regular sparse-PCA approaches, we take a penalised likelihood estimation approach, and noting that the likelihood is incomplete, we suggest a novel EM algorithm to perform estimation. The algorithms developed are computationally efficient, and give users a new method for imposing weakly informative (sparse) structural priors on the factor model. The data-driven estimation of the loadings support contrasts with the hard constraints that are more traditional in the use of DFMs.

The analysis within this paper is empirical in nature, we consider three aspects: a) how our EM algorithm performs in recovering the true sparsity pattern in the factor loadings; b) how the model contrasts with alternative models in a predictive setting, e.g.~where we want to forecast either all the $p$ time series, or just a subset of these; and c) how the model and estimation routine can be used in practice to extract insights from complex real-world datasets. The first two points are illustrated through extensive synthetic experiments, whilst for the latter, we give an example application to a set of smart-meter data from across our University campus. To our knowledge this is the first time a DFM has been used to study building level energy data, and illustrates some of the benefits that come from imposing sparsity in terms of increasing the interpretability of the model.

\section{Background and Related Work}

Canonically, the dynamics of the latent factors in the DFM are specified as a stationary VAR(1) model:
\begin{equation}
    \label{eq:state}
    \bF_t = \bA\bF_{t-1} + \bu_t \, ,
\end{equation}
where $\bu_t$ is a zero-mean series of disturbances with covariance matrix $\bsig_u$. Furthermore, the idiosyncratic errors, $\beps_t$, in (\ref{eq:measurement}), are commonly assumed to be zero-mean and cross-sectionally uncorrelated, meaning their covariance matrix, which we denote $\bsig_\epsilon$, is diagonal---models with these assumptions are termed \emph{exact}. However, as shown by \cite{doz2011two}, even when this assumption is relaxed and the idiosyncratic errors are (weakly) cross correlated, referred to as an \emph{approximate} DFM, then consistent estimation of the factors is possible as $(n,p) \rightarrow \infty$. Therefore, the `curse of dimensionality', often a burden for analysing time-series models, can actually be beneficial in DFMs. 

\subsection*{Estimation}

The measurement equation (\ref{eq:measurement}) along with the state equation (\ref{eq:state}) form a state space model. A simple approach to estimate factor loadings is to consider the first $r$ eigenvectors of the sample covariance matrix of $\bX$, essentially applying PCA to the time series. This has been extensively reviewed in the literature \citep{stock2002forecasting, bai2003inferential, doz2020dynamic}. When mild conditions 
are placed on the correlation structure of idiosyncratic errors, the PCA estimator is the optimal 
non-parametric\footnote{In the sense that temporal dependence is not restricted to that encoded via a parametric model} estimator for a large approximate DFM. With even tighter conditions of spherical idiosyncratic components, i.e.~they are i.i.d.~Gaussian, then the PCA estimator is equivalent to the maximum likelihood estimator \citep{doz2020dynamic}. The problem with using non-parametric PCA methods to estimate the loading structure is that there is no consideration of the dynamics of the factors or idiosyncratic components. In particular, there is no feedback from the estimation of the state equation (\ref{eq:state}) to the measurement equation (\ref{eq:measurement}). For this reason, it is preferable to use parametric methods that are able to account for temporal dependencies in the system.

An alternative approach is proposed in \citep{giannone2008nowcasting} whereby the initial estimates of the factors and loadings are derived from PCA, the VAR(1) parameters are estimated from these preliminary factors, before updating the factor estimates using Kalman smoothing. This two-stage approach has been theoretically analysed in \cite{doz2011two} and successfully applied to the field of nowcasting in many national statistical institutes and central banks. 
%
%
%
The Kalman smoothing step in particular is very helpful for handling missing data, whether it be backcasting missing at the start of the sample, forecasting missing data at the end of the sample\footnote{Missing data at the end of the sample, commonly referred to as the `ragged edge' problem, is very common in macroeconomic nowcasting applications. It is caused by time series used in the model having differing publication delays, and hence forming a ragged edge of missingness at the end of sample.} or interpolating arbitrary patterns of missing data throughout the sample.

\cite{banbura2014maximum} build on the DFM representation of \cite{watson1983alternative} and adopt an expectation-maximisation (EM) algorithm to estimate the system (\ref{eq:measurement})-(\ref{eq:state}) with a quasi maximum likelihood estimation (QMLE) approach. 
%
%
\cite{doz2012quasi}, \citet{Bai2016}, and \cite{barigozzi2022quasi} provide theoretical results whereby, as $(n,p) \rightarrow \infty$, the QMLE estimates (based on an exact Gaussian DFM) are consistent under milder assumptions allowing for correlated idiosyncratic errors. The EM approach to estimation is beneficial as it allows feedback between the estimation of the factors and the loadings, and thus handle arbitrary patterns of missing data.

\subsection*{Relation to Current Work}




In the literature, the idea of a sparse DFM is not new. A classic approach is to use factor rotations that aim to minimise the complexity in the factor loadings to make the structure simpler to interpret. See \cite{kaiser1958varimax} for the well-established varimax rotation and see \cite{carroll1953analytical} and \cite{jennrich1966rotation} for the well-established quartimin rotation. For a recent discussion paper on the varimax rotation see \cite{rohe2020vintage}. An alternative approach based on LASSO regularisation is to use sparse principle components analysis (SPCA) \citep{zou2006sparse} in place of regular PCA on the sample covariance matrix in the preliminary estimation of factors and loadings, i.e.~in stage one of the two-stage approach by \cite{giannone2008nowcasting}. For factor modelling, it has been used by \cite{croux2011sparse} in a typical macroeconomic forecasting setting where they consider a robustified version. \cite{kristensen2017diffusion} use SPCA to estimate diffusion indexes with sparse loadings. \cite{despois2022identifying} prove that SPCA consistently estimates the factors in an approximate factor model if the $\ell_1$ penalty is of $\mathcal{O}(1/\sqrt{p})$. They also compare SPCA with factor rotation methods and show an improved performance when the true loadings structure is sparse. 

Unlike previous research, our methodology implements regularisation within an EM algorithm framework, allowing us to robustly handle arbitrary patterns of missing data, model temporal dependence in the processes, and impose weakly informative (sparse) prior knowledge on the factor loadings. We argue that in settings where autocorrelation is moderately persistent, that the feedback provided through our EM procedure is important in aiding recovery of the factor loadings, as well as producing accurate forecasts.

The rest of the paper is structured as follows. In Section \ref{sec:3} we formalise our DFM model and the sparsity assumptions placed on the loading matrices. Sect.~\ref{sec:4} presents a regularised likelihood estimator for the model parameters, and introduces an EM algorithm to enable finding feasible estimates. Sect.~\ref{sec:5} discusses how we implement the method using the R package \verb|sparseDFM| \citep{mosleyJSS}. Numerical results, including simulation studies and real data analysis, are presented in Sects.~\ref{sec:6} and \ref{sec:7}, respectively. The paper concludes with a discussion of the results, and how the models and estimators can be further generalised to provide flexibility to users.


\section{The Sparse DFM}\label{sec:3}

Consider the $p$-variate time series $\{\bX_t\}$ and $r$ factors $\{\bF_t\}$ related according to the model
\begin{align}
    \bX_t &= \blam_0 \bF_t + \beps_t \, \label{eq:M_orig} \\
    \bF_t &= \bA\bF_{t-1} + \bu_t \, \, ,\nonumber
\end{align}
where $\{\beps_t\}$ and $\{\bu_t\}$ are multivariate white noise processes. For simplicity we assume $E[\beps_t \beps_t^\top]=\Sigma_{\epsilon} = \mathrm{diag}(\bsigma^2_{\epsilon})$ and $\bsigma^2_{\epsilon}\in\mathbb{R}_+^p$ is a vector of idiosyncratic variances. Similarly, let $E[\bu_t \bu_t^\top]=\Sigma_{u}$ and assume the eigenvalues of the VAR matrix are bounded $\|\bA\| <1$, thus the latent process is assumed stationary. This model corresponds to an exact DFM, where all the temporal dependence is modelled via the latent factors.

In this context, our notion of sparsity relates to the assumption that many of the entries in $\blam_0$ will be zero. For instance, let the support of the $k$th column of the loading matrix be denoted 
\[
\mathcal{S}_k := \mathrm{supp}(\Lambda_{0;\cdot,k})\subseteq \{1,\ldots,p\}\;,
\]
such that $s_k=\vert\mathcal{S}_k\vert$. We refer to a DFM as being sparse if $s_k<p$ for some or all of the $k=1,\ldots,r$ factors. In practice, this is an assumption that many of the observed series are driven by only a few ($r$) latent factors, and that for many series only a subset of the factors will be relevant.

\subsection{Consistency and Pervasiveness}
 
In the sparse situation, whereby $s_k< p$, we will be able to model only a subset of the observations with each factor. To enable us to model all $p$ variables and gain information relating to the $r$ factors as $n,p$ increase we assume a couple of conditions on the specification. First, that the support of the observations, and the union of factor supports is equal, i.e.~$\cup_{k=1}^r\mathcal{S}_{k}=\{1,\ldots,p\}$, thus all observations are related to at least one of the factors. Second, that the support for each factor grows with the number of observed variables, in that $\{s_k\}$ is a non-decreasing sequence in $p$. Assumptions of this form would allow us, in principle, to assess the consistency of factor estimation as $p$ grows.

This asymptotic analysis in $p$ (and $n$) contrasts with the traditional setting with a fixed $p$---for which the factors cannot be consistently recovered and can only be approximated, with error that depends on the signal-noise-ratio $\|\blam_0\Sigma_F\blam_0^\top]\|/\|\Sigma_\epsilon\|$, where $\Sigma_F=E[\bF_t \bF_t^\top]$ \citep{Bai2016}. Intuitively, this is due to the fact that if $p$ is fixed, then we cannot learn anything more about the factor at a specific time $t$, as we do not get more information on the factors as $n$ increases, instead we just get more samples (at different time points) relating to the series $\{\bF_t\}$. When we go to the doubly asymptotic, or just $p\rightarrow \infty$ setting, then if the number of factors $r$ is fixed or restricted to slowly grow in $n$ then we can not only recover structures relating to $\{\bF_t\}$, e.g.~the specification of $\bA$, but we can also get more information relating to the factor at the specific time $t$ \citep{Bai2016,barigozzi2022quasi}. One way to ensure this growing information about the factors is to assume that they are in some sense pervasive---the more variables $p$ we sample, the more this tells us about the $r$ factors. We note, that for a more formal analysis of the DFM, a usual pervasiveness assumption placed on the loading conditions is given by \citet{doz2011two}, whereby $\lim_{p\rightarrow\infty} p^{-1}\lambda_{\min}(\blam_0^{\top}\blam_0)> 0$, i.e. the average loading onto the least-influential factor is bounded away from zero. 

In this paper, we choose to focus on the empirical performance of our estimator, thus we do not formalise the sparsity assumptions further. However, it is worth noting our empirical studies meet the pervasiveness assumptions regarding the support of the factor loadings.


\subsection{Identifiability}

In the following section, we will consider a QMLE estimator for the factor model based on assuming Gaussian errors $\beps_t$ and $\bu_t$, it is thus of interest to consider how the associated likelihood relates to the factors and their loadings. Adopting a Gaussian error structure and taking expectations over the factors, the likelihood for (\ref{eq:M_orig}) is given by
\begin{align*}
\mathcal{L}(\blam) &\propto \log\det(\blam^{\top} \bsig_F \blam + \bsig_{\epsilon}) \\ &- \frac{1}{2}\mathrm{tr}\left[(\blam^{\top} \bsig_F\blam + \bsig_{\epsilon})^{-1}\frac{1}{n}\sum_{t=1}^n \bX_t \bX_t^{\top}\right]\;.
\end{align*}
%

An obvious identifiability issue arises here, such that if $\tilde{\blam}=\blam Q$, $\tilde{\bF}_t=Q \bF_t$, for any unitary matrix $Q^{\top}=Q^{-1}$, we have $\mathcal{L}(\tilde{\blam})=\mathcal{L}(\blam)$. Now consider the case of $\tilde{\blam}_0$, i.e.~performing a rotation on the true loadings, denote the set of all possible equivalent loading as
\begin{equation}
\mathcal{E} := \{\blam_0^*Q \;\vert\;Q^{\top}=Q^{-1}\;,\;Q\in\mathbb{R}^{r\times r}\}\;.
\end{equation}

The invariance of the likelihood to elements of this set mandates that theoretical analysis of the DFM is typically constructed in a specific frame of reference, c.f.~\citet{doz2011two,doz2012quasi,Bai2016}.
Interestingly, our sparsity assumptions restrict the nature of this equivalence class considerably, in that only loading matrices with sparse structure are permitted. In general, there will still be multiple sparse representations that are allowed, and the issue of the scale invariance remains, however, the latter can be fixed by imposing a further constraint on the norms of the loading matrices. In this work, we demonstrate empirically that it is possible to construct estimators that are consistent up to rotations that maintain an optimal level of sparsity, in the sense that the true loading matrix is given by
\begin{equation}
\blam_0 \in \arg\min_{\blam\in \mathcal{E}}\sum_{k=1}^r\|\blam_{\cdot,k}\|_0\;.\label{eq:maximal_sparsity}
\end{equation}
where $\|\blam_{\cdot,k}\|_0:=\vert\mathrm{supp}(\blam_{\cdot,k})\vert$ counts the number of non-zero loadings. More generally (see Remark \ref{remark:invariance}) we could consider selecting on the basis of the $\ell_q$ norm, $\|\blam\|_q:=(\sum_{ik}\Lambda_{i,k}^q)^{1/q}$, the $\ell_1$ norm may still provide selection,  however, the $\ell_2$ norm provides no selection as it maintains the rotational invariance of the likelihood.
In this paper, we restrict our equivalence set on the basis of the $\ell_0$ norm, as above, that is, we specify the true loading matrices as those that maintain the highest number of zero values after consideration for all unitary linear transformations.

In practice, these issues mean we are unable to recover the correct sign of the factor loadings, whilst columns in the loading matrix may also be permuted, e.g.~factor $k$ can be swapped (under permutation of the columns in the loading matrix) with factor $l$, for any $k,l\in\{1,\ldots,r\}$. These are the same identifiability issues which we face in PCA, whereby the eigenvectors can be exchanged in terms of order and direction.

\begin{figure}
\centering\includegraphics[width=0.7\linewidth]{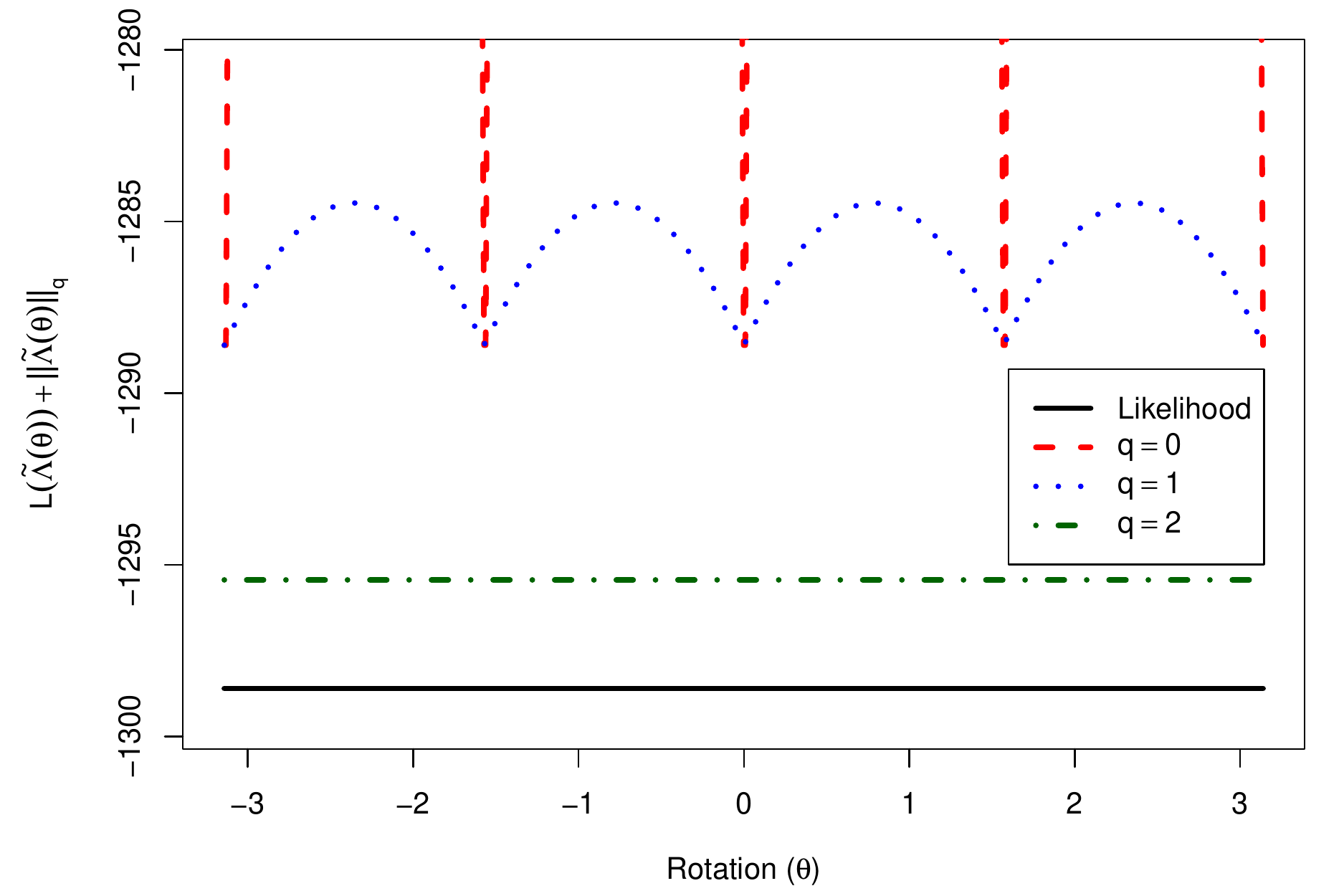}
\caption{The impact of rotation on the function $\mathcal{L}(\tilde{\blam}(\theta))+\|\tilde{\blam}(\theta)\|_q$, in the case of $q=0,1$ the set of feasible $\blam_0$ from (\ref{eq:maximal_sparsity}) is restricted to the points $\theta\in\{0,\pm \frac{1}{2}\pi,\pm \pi\}$ corresponding to either swapping columns, or flipping signs.}\label{fig:rotation}
\end{figure}

\begin{remark}{Sparsity and Invariance \label{remark:invariance}}
To illustrate how the sparsity constraint (\ref{eq:maximal_sparsity}) breaks the more general invariance that regular DFMs suffer, we can consider the  quantity $\|\blam_0^* Q_{\mathrm{rot}}(\theta)\|_q$, where $Q_{\mathrm{rot}}(\theta)\in\mathbb{R}^{2\times 2}$ is a rotation matrix with argument $\theta\in(-\pi,\pi)$, and $\blam_0^*\in\mathbb{R}^{10\times 2}$ has the first column half filled with ones, and the rest zero, the second column is set to be one minus the first.
%
%
As we see from Fig.~\ref{fig:rotation}, without the additional restriction on our specification of $\blam_0$, via Eq.~\ref{eq:maximal_sparsity}, we would not be able to determine a preference for any particular element from the set $\mathcal{E}:=\{\blam_0^* Q_{\mathrm{rot}}(\theta) :\vert\;\theta\in\{-\pi,\pi\}\}$. 
\end{remark}

\section{Estimation}\label{sec:4}


Under the Gaussian error assumption, and collecting all parameters of the DFM (\ref{eq:M_orig}) in $\btheta = (\blam, \bA, \bsig_{\beps}, \bsig_{\bu})$, we are able to write the joint log-likelihood of the data $\bX_t$ and the factors $\bF_t$ as:
\begin{align}
     &\log\mathcal{L}(\bX,\bF;\btheta) \label{eq:logl}\\
     & = -\frac{1}{2}\log \vert\mathcal{\bP}_0\vert - \frac{1}{2}(\bF_0 - \bm{\alpha}_0)^\top \mathcal{\bP}_0^{-1} (\bF_0 - \bm{\alpha}_0) \nonumber \\
     & - \frac{n}{2} \log \vert\bsig_{\bu}\vert - \frac{1}{2}\sum_{t=1}^n \bu_t^\top \bsig_{\bu}^{-1} \bu_t \nonumber \\
     & - \frac{n}{2} \log \vert\bsig_{\beps}\vert - \frac{1}{2}\sum_{t=1}^n \beps_t^\top \bsig_{\beps}^{-1} \beps_t \, \nonumber
\end{align}
where $\beps_t = \bX_t-\blam_0 \bF_t$, $\bu_t=\bF_t-\bA\bF_{t-1}$, and we have assumed an initial distribution at $t=0$ of the factors as $\bF_0 \sim N(\bm{\alpha}_0, \mathcal{\bP}_0)$. 

We propose to induce sparsity in our estimates using the familiar $\ell_1$ penalty, with motivation similar to that of the LASSO \citep{Tibshirani1996}. Alternative penalty functions are available, however, the convexity of the $\ell_1$ penalty is appealing. Even though the overall objective for the parameters is non-convex, due to the rotational invariance of the log-likelihood, the convexity of the penalty ensures we can quickly and reliably apply the sparsity constraints. We will make use of this structure in the algorithms we construct to find estimates in practice.
It is worth noting that our focus here is on the factor loadings, and thus this is the object we regularise, possible extensions could consider additional/alternative constraints, for instance on the latent VAR matrix. 

Our proposed estimator attempts to minimise a penalised negative log-likelihood, as follows
\begin{align}
\hat{\btheta} = \arg\min_{\btheta} -\log\mathcal{L}(\bX,\bF;\btheta) + \alpha R(\blam)\;,\label{eq:reg_lik}
\end{align}
where $\alpha\ge 0$. A larger $\alpha$ corresponds to a higher degree of shrinkage on the loadings, e.g.~for a larger $\alpha$ we would expect more zero values in the loading matrices. 

\subsection{A Regularised Expectation Maximisation Algorithm}
The regularised likelihood (\ref{eq:reg_lik}) is incomplete, as whilst we have observations, we do not observe the factors. To solve this problem, we propose to construct an Expectation-Maximisation (EM) framework where we take expectations over the factors (fixing the parameters), then conditional on the expected factors we maximise the log-likelihood with respect to the parameters $\btheta$, we iterate this process until our estimates converge.

The EM algorithm involves calculating and maximising the expected log-likelihood of the DFM conditional on the available information $\bomega_n$. Given the log-likelihood in (\ref{eq:logl}), the conditional expected log-likelihood is 
\begin{align}
     &\E \left[\log\mathcal{L}(\bX,\bF;\btheta)\vert\bomega_n\right]  = -\frac{1}{2}\log \vert\mathcal{\bP}_0\vert \label{eq:clogl}\\
     &  - \mathrm{tr}\left\{\mathcal{\bP}_0^{-1} \E \left[(\bF_0 - \bm{\alpha}_0)(\bF_0 - \bm{\alpha}_0)^\top \vert\bomega_n\right] \right\} \nonumber \\
     & - \frac{n}{2} \log \vert\bsig_{\bu}\vert - \frac{1}{2}\sum_{t=1}^n \mathrm{tr}\left\{ \bsig_{\bu}^{-1} \E \left[\bu_t^\top \bu_t \vert \bomega_n\right]\right\} \nonumber \\
     & - \frac{n}{2} \log \vert\bsig_{\beps}\vert - \frac{1}{2}\sum_{t=1}^n \mathrm{tr}\left\{ \bsig_{\beps}^{-1} \E\left[\beps_t^\top\beps_t \vert \bomega_n\right] \right\} \; . \nonumber
\end{align}
Ultimately, we wish to impose our regularisation on the expected log-likelihood, our feasible estimator being given by
\begin{align}
\hat{\btheta} = \arg\min_{\btheta} \left[-\E \left[\log\mathcal{L}(\bX,\bF;\btheta)\vert\bomega_n\right] + \alpha \|\blam\|_1\right] \;. \label{eq:reg_elik}
\end{align}

\subsubsection{Maximisation-Step}

We use the following notation for the conditional mean and covariances of the state:
\begin{align*}
    \ba_{t\vert s} &= \E[\bF_t\vert\bomega_s] \, ,\\
    \bP_{t\vert s} &= \mathrm{Cov}[\bF_t\vert\bomega_s] \, ,\\
    \bP_{t,t-1\vert s} &= \mathrm{Cov}[\bF_t, \bF_{t-1}\vert\bomega_s] \, .
\end{align*}
conditional on all information we have observed up to a time $s$, denoted by $\bomega_s$.

As shown in \cite{banbura2014maximum}, the maximisation of (\ref{eq:clogl}) results in the following expressions for the parameter estimates: 
\begin{equation}
    \hat{\bm{\alpha}}_0 = \ba_{t\vert n}\quad;\quad \hat{\mathcal{\bP}}_0 =\bP_{t\vert n}    \label{eq:init} 
\end{equation}
and letting $\bS_{t\vert n} = \ba_{t\vert n}\ba_{t\vert n}^{\top} + \bP_{t\vert n}$, and $\bS_{t,t-1\vert n} = \ba_{t\vert n}\ba_{t-1\vert n}^{\top} + \bP_{t,t-1\vert n}$ we have
\begin{align}
    \hat{\bA} &= \left(\sum_{t=1}^n \bS_{t-1\vert n}\right)^{-1} \left(\sum_{t=1}^n \bS_{t,t-1\vert n}\right)\label{eq:estA} \, , \\
    \hat{\bsig}_{\bu} &= \frac{1}{n} \sum_{t=1}^n\left[\bS_{t\vert n} - \hat{\bA} \left( \bS_{t-1,t\vert n}  \right) \right] \, . \label{sigu}
\end{align}

To minimise (\ref{eq:reg_elik}) for parameters $\blam$ and $\bsig_{\beps}$, we should also consider there might be missing data in $\bX_t$. Let us define a selection matrix $\bW_t$ to be a diagonal matrix such that 
\begin{equation*}
W_{t,ii}=\begin{cases}
1 & \mathrm{\mathrm{if}\;}X_{i,t}\;\mathrm{observed}\\
0 & \mathrm{\mathrm{if}\;}X_{i,t}\;\mathrm{missing}
\end{cases}
\end{equation*}
and note that $\bX_t = \bW_t\bX_t + (\bI-\bW_t)\bX_t $. The update for the idiosyncratic error covariance is then given by
\begin{align}
\hat{\bsig}_{\beps} &= \frac{1}{n}\sum_{t=1}^n \mathrm{diag}\Bigg[ \bW_t \bigg( \bX_t\bX_t^\top\ - 2\bX_t \ba_{t\vert n}^\top \hat{\blam}^\top \nonumber \\
     \qquad \qquad
     & + \hat{\blam}\bS_{t\vert n}\hat{\blam}^\top\bigg) + (\bI - \bW_t)\hat{\bsig}_{\beps}^*(\bI - \bW_t) \Bigg] \, , \label{eq:sigeps}
\end{align}
where $\hat{\bsig}_{\beps}^*$ is obtained from the previous EM iteration. As noted in Algorithm 1, in practice we update $\hat{\bsig}_{\beps}$ after estimating $\hat{\blam}$, as the former is based on the difference between the observations and the estimated common component. The following section details precisely how we practically obtain sparse estimates for the factor loadings, the estimates can then be used in (\ref{eq:sigeps}) and thus complete the M-step of the algorithm.

\subsubsection{Incorporating Sparsity}

In this work, we propose to update $\hat{\blam}$ by constructing and Alternative Directed Method of Moments (ADMM) algoritghm \citep{boyd2011distributed} to solve (\ref{eq:reg_elik}) with the parameters ($\hat{\bA},\hat{\bsig}_u,\hat{\bm{\alpha}}_0,\hat{\mathcal{\bP}}_0$) fixed. The algorithm proceeds by sequentially minimising the augmented Lagrangian 
\begin{align}
    \mathcal{C}(\blam, \bZ, \bU) &:=-\E\left[\log\mathcal{L}(\bX,\bF;\btheta)\vert\bomega_n\right]\label{eq:augmented}\\ 
    & \quad+\alpha\|\bZ\|_{1}+\frac{\nu}{2}\|\blam-\bZ+\bU\|_{F}^{2} \, ,\nonumber
\end{align}
where $\bZ\in\mathbb{R}^{p\times r}$ is an auxiliary variable, $\bU\in\mathbb{R}^{p\times r}$
are the (scaled) Lagrange multipliers and $\nu$ is the scaling term. Under equality conditions relating the auxilary ($\bZ$) to the primal ($\blam$) variables, this is equivalent to minimising (\ref{eq:reg_elik}), e.g.
\begin{align*}
&\arg\min_{\bZ=\blam} \max_{\bU}\mathcal{C}(\blam, \bZ, \bU) \\
&= \arg\min_{\blam}\left[-\E \left[\log\mathcal{L}(\bX,\bF;\btheta)\vert\bomega_n\right] + \alpha \|\blam\|_1\right]
\end{align*}
as (\ref{eq:reg_elik}) is convex in the argument $\blam$ with all other parameters fixed, this argument holds for any $\nu>0$ \citep{boyd2011distributed,Lin2014}.

The augmented Lagrangian (\ref{eq:augmented}) can be sequentially minimised via the following updates\footnote{For the full derivation of $\bZ$ and $\bU$ refer to \cite{boyd2011distributed}. For the full derivation of $\blam$ refer to the software paper implementing this algorithm of \cite{mosleyJSS}.
}
\begin{align*}
\blam^{(k+1)} & =\argmin_{\blam}\mathcal{C}(\blam,\bZ^{(k)},\bU^{(k)})\\
\bZ^{(k+1)} & =\argmin_{\bZ}\mathcal{C}(\blam^{(k+1)},\bZ,\bU^{(k)})\\
&=\mathrm{soft}(\blam^{(k+1)}+\bU^{(k)};\alpha/\nu)\\
\bU^{(k+1)} & =\bU^{(k)}+\blam^{(k+1)}-\bZ^{(k+1)} \, .
\end{align*}
for $k=0,1,2,\ldots,$ until convergence.
The first (primal) update is simply a least-squares type problem, whereby on vectorising $\blam$ one finds
\begin{align}
\mathrm{vec}(\blam^{(k+1)}) = &\left( \sum_{t=1}^n \bS_{t\vert n} \otimes \bW_t\bsig_{\beps}^{-1}\bW_t + \nu\bI_{pr} \right)^{-1}  \nonumber \\
&\mathrm{vec}\bigg[ \sum_{t=1}^n\bW_t\bsig_{\beps}^{-1}\bW_t\bX_t \ba_{t\vert n}^\top \nonumber\\
&+  \nu(\bZ^{(k)}-\bU^{(k)})\bigg] \, .
\label{eq:Lambda_sparse}
\end{align}

\begin{remark}[Exploiting Dimensionality Reduction]
For the $\blam^{(k+1)}$ update, the dimensionality of the problem is quite large, leading to a na\"ive per-iteration cost of order $\mathcal{O}(r^3p^3)$. A more efficient method for this step can be sought by looking at the specific structure of the matrix to be inverted. Define $\mathcal{A}_t = \bS_{t \vert n}$, $\mathcal{B}_t = \bW_t\bsig_{\beps}^{-1}\bW_t$, and $\mathcal{C} = \sum_{t=1}^n\bW_t\bsig_{\beps}^{-1}\bW_t\bX_t\ba_{t\vert n}^{\top} + \nu(\bZ^{(k)}-\bU^{(k)})$, then the solution (\ref{eq:Lambda_sparse}) can be written as 
\begin{align*}
    \mathrm{vec}(\blam) &= \left( \sum_{t=1}^n\mathcal{A}_t \otimes \mathcal{B}_t + \nu\bI_{pr} \right)^{-1} \mathrm{vec}(\mathcal{C}) \\
    &= \mathcal{D}^{-1}\mathrm{vec}(C) \, . 
\end{align*}
Since $\bsig_{\beps}$ is diagonal in an exact DFM, $\mathcal{B}_t$ is also diagonal and thus $\mathcal{D}$ is made up of $r^2$ blocks such that each $(i,j)^{th}$ block is a diagonal matrix of length $p$ for $i,j=1,\dots,r$. To speed up the computation, we note that $\nu\bI_{pr} = \nu\bI_r \otimes \bI_p$ and use the properties of commutation matrices \citep[p.~54]{magnus2019matrix},
denoted by $\bK_{rp}$, to write 
\begin{align}
    &\left( \sum_{t=1}^n\mathcal{A}_t \otimes \mathcal{B}_t +  \nu\bI_r \otimes \bI_p\right)^{-1}  \nonumber \\
    &= \left[ \sum_{t=1}^n \bK_{rp}(\mathcal{B}_t \otimes \mathcal{A}_t) \bK_{pr} + \bK_{rp}(\bI_p \otimes \nu\bI_r)\bK_{pr} \right]^{-1} \nonumber \\
    &= \bK_{rp}   \left(\sum_{t=1}^n (\mathcal{B}_t \otimes \mathcal{A}_t) + (\bI_p \otimes \nu\bI_r)\right)^{-1} \bK_{pr}  \, .
    \label{eq:fast_inv}
\end{align}
The matrix needing to be inverted in the final line of equation (\ref{eq:fast_inv}) is now a block diagonal matrix. We can extract each of the $1,\ldots,p$ blocks separately and invert them one-by-one. The final result from (\ref{eq:fast_inv}) has the expected block structure with a diagonal matrix in each block but we can stack them into a cube to save storage. Overall, the operations can be completed with cost $\mathcal{O}(r^3p)$. Given that this needs to be performed for every iteration of the EM algorithm, our commutation trick results in significant computational gains.
\end{remark}

Whilst other optimisation routines could be used to estimate the sparse loadings, the ADMM approach is appealing as it allows us to split (\ref{eq:reg_elik}) into sub-problems that can easily be solved. If one wished to incorporate more specific/structured prior knowledge, this approach can easily be altered to impose these assumptions, for instance, future work could consider group-structured regularisation allowing for more informative prior knowledge on the factor loadings to be incorporated. Hard constraints, e.g.~where we require a loading to be exactly zero can also be incorporated at the $\bZ$ update stage by explicitly setting some entries to be zero.

\subsubsection{Expectation Step}

So far, we have discussed how to update the parameters conditional on the quantities $\E[\bF_t\vert\bomega_n]$ $\mathrm{Cov}[\bF_t\vert\bomega_n]$, and $\mathrm{Cov}[\bF_t, \bF_{t-1}\vert\bomega_n]$. In our application, under the Gaussian error assumption, these expectations can be easily calculated via the Kalman smoother. For completeness, we detail this step in the context of the DFM model, as well as discussing some methods to speed up the computation which make use of the exact DFM structure.

The classical multivariate Kalman smoother equations can be slow when $p$ is large. However, since we assume $\bsig_{\beps}$ is diagonal, we can equivalently filter the observations $\bX_t$ one element at a time, as opposed to updating all $p$ of them together as in the classic approach \citep{durbin2012time}. As matrix inversion becomes scalar divisions, huge speedups are possible. This approach, sometimes referred to as the univariate treatment, sequentially updates across both time and variable index when filtering/smoothing. Let us define the individual elements $\bX_t = (X_{t,1},\dots,X_{t,p})^\top$, $\blam = (\blam_1^\top,\dots,\blam_p^\top)^\top$, $\bsig_{\beps} = \mathrm{diag}(\sigma_{\epsilon 1}^2,\dots,\sigma_{\epsilon p}^2)$. Following \citet{koopman2000fast} we expand the conditional expectations according to
\begin{align*}
\ba_{t,i} &= \E[\bF_{t}\vert\bomega_{t-1},X_{t,1}, \dots, X_{t,i-1}] \, , \\
\ba_{t,1} &= \E[\bF_{t}\vert\bomega_{t-1}] \, , \\
\bP_{t,i} &= \mathrm{Var}[\bF_{t}\vert\bomega_{t-1},X_{t,1}, \dots, X_{t,i-1}] \, , \\
\bP_{t,1} &= \mathrm{Var}[\bF_{t}\vert\bomega_{t-1}] \, ,
\end{align*}
for $i=1,\dots,p$ and $t=1,\dots,n$. The univariate treatment now filters this series over indices $i$ and $t$. This is equivalent in form to the multivariate updates of the classic \citep{shumway1982approach} approach, except that the $t$ subscript now becomes a $t,i$ subscript, and the $t\vert t$ subscript now becomes $t,i+1$. 
\begin{align*}
v_{t,i} &= X_{t,i}-\blam_{i}\ba_{t,i}, \\
C_{t,i} &= \blam_{i}\bP_{t,i}\blam_{i}^\top+\sigma^2_{\epsilon,i}, \\
\bK_{t,i} &= \bP_{t,i}\blam_{i}^\top C_{t,i}^{-1}, \\
\ba_{t,i+1} &= \ba_{t,i}+\bK_{t,i}v_{t,i}, \\
\bP_{t,i+1} &= \bP_{t,i}-\bK_{t,i}C_{t,i}\bK_{t,i}^\top,
\end{align*}
for $i=1,\ldots,p$ and $t=1,\ldots,n$. If $X_{t,i}$ is missing or $C_{t,i}$ is zero, we omit the term containing $\bK_{t,i}$. The transition to $t+1$ is given by the following prediction equations:
\begin{align*}
\ba_{t+1,1} &= \bA \ba_{t,p+1}, \\
\bP_{t+1,1} &= \bA \bP_{t,p+1}\bA^\top+\bsig_{\bu}.
\end{align*}
These prediction equations are exactly the same as the multivariate ones (i.e., predictions are not treated sequentially but all at once). From our perspective, this univariate treatment may be more appropriately referred to as performing univariate updates plus multivariate predictions.

Unlike \cite{shumway1982approach}, the measurement update comes before the transition; however, we can revert to doing the transition first if our initial state means and covariances start from $t=0$ instead of $t=1$. Likewise, univariate smoothing is defined by:
\begin{align*}
\bL_{t,i} &= \bI_m - \bK_{t,i} \blam_{i}, \\
\bb_{t,i-1} &= \blam_{i}^\top C_{t,i}^{-1}v_{t,i}+\bL_{t,i}^\top \bb_{t,i}, \\
\bJ_{t,i-1} &= \blam_{i}^\top C_{t,i}^{-1}\blam_{i}+\bL_{t,i}^\top\bJ_{t,i}\bL_{t,i}, \\
\bb_{t-1,p} &= \bA^\top \bb_{t,0}, \\
\bJ_{t-1,p} &= \bA^\top \bJ_{t,0}\bA,
\end{align*}
for $i=p,\ldots,1$ and $t=n,\ldots,1$, with $\bb_{n,p}$ and $\bJ_{n,p}$ initialised to $0$. Again, if $X_{t,i}$ is missing or $C_{t,i}$ is zero, drop the terms containing $\bK_{t,i}$. Finally, the equations for $\ba_{t\vert n}$ and $\bP_{t\vert n}$ are:
\begin{align*}
\ba_{t\vert n} &= \ba_{t,1}+\bP_{t,1}\bb_{t,0}, \\
\bP_{t\vert n} &= \bP_{t,1}-\bP_{t,1}\bJ_{t,0}\bP_{t,1}.
\end{align*}
These results will be equivalent to $\ba_{t\vert n}$ and $\bP_{t\vert n}$ from the classic multivariate approach, yet obtained with substantial improvement in computational efficiency. 
In order to calculate the cross-covariance matrix $\bP_{t,t-1\vert n}$, we use \cite{jong1988covariances}'s theorem:
\begin{equation}\label{eq:deJong}
\bP_{t,t-1\vert n} = \bP_{t\vert n}(\bP_{t\vert t-1})^{-1}\bA \bP_{t-1\vert t-1}.
\end{equation}

\subsection{Parameter Tuning}\label{sec:tuning}

There are two key parameters that need to be set for the DFM model. The first is to select the number of factors, and the second is to select an appropriate level of sparsity. One may argue that these quantities should be selected jointly, however, in the interests of computational feasibility, we here propose to use heuristics, first selecting the number of factors, and then deciding on the level of sparsity. This mirrors how practitioners would typically apply the DFM model, where there is often a prior for the number of relevant factors (or more usually an upper bound). Both the number of factors, and the structure of the factor loadings impact the practical interpretation of the estimated factors.

\subsubsection{Choosing the Number of Factors}
To calculate the number of factors to use in the model we opt to take the information criteria approach of \cite{bai2002determining}. There are several criteria that are discussed in the literature, for example, the paper of \citet{bai2002determining} suggests three forms\footnote{In our experiments and applications, we compared all criteria and they typically give similar results within $\pm1$ of each other, for simplicity, only one IC is presented here.}. For this paper, we use the criteria of the following form:
\begin{equation}
    IC(r) = \log V_r(\bar{\bF},\bar{\blam}) + r \left( \frac{n+p}{np} \right)\log \min(n,p)\;, \label{eq:IC2}
\end{equation}
where 
\[
V_r(\bar{\bF},\bar{\blam}) = \frac{1}{np}\sum_{i=1}^p\sum_{t=1}^n\E[\bar{\epsilon}_{i,t}^2]
\]
and $\bar{\epsilon}_{i,t} = X_{t,i}-\bar{\blam}_{i,\cdot}\bar{\bF}_t$ is found using PCA when applied to the standardized data. The preliminary factors $\bar{\bF}$ correspond to the principle components, and the estimated loadings $\bar{\blam}$ corresponding to the eigenvectors. Should the data contain missing values, we first interpolate the missing values using the median of the series and then smooth these with a simple moving window.
\begin{remark} We note that ideally one may wish to apply the EM procedure to get more refined estimates of both the factors and loadings, however, in the interests of computational cost and in-line with current practice we propose to use the quick (prelininary) estimates above, denoted with $\bar{\blam}$ rather than $\hat{\blam}$.
\end{remark}

\subsubsection{Tuning the Regulariser}

Once a number of factors $r$ has been decided, we tune $\alpha$ by performing a simple search over a logarithmically spaced grid and minimise a Bayesian Information Criteria defined as
\begin{equation}
    BIC(\alpha) = \log\left( V_{\alpha}(\hat{\bF},\hat{\blam}) \right) + \frac{\log(np)}{np}\sum_{k=1}^r \hat{s}_k  \, ,\label{eq:BIC}
\end{equation}
where $\hat{s}_k$ is the number of non-zero entries in the $k$th column of the estimated loading matrix. In this case, we run the EM algorithm until convergence (usually after a dozen or so iterations) and then evaluate the BIC using the resulting $\hat{\bF}$ and $\hat{\blam}$, this procedure is repeated for each $\alpha$ in the grid. An example of the resulting curve can be seen in the empirical application of Section \ref{sec:7}. To limit searching over non-optimal values, an upper limit for $\alpha$ is set whereby, if the loadings for a particular factor are all set to zero, then we terminate the search.

\begin{remark}Tuning both the number of factors, and the regulariser for these models is a topic of open research and discussion. Indeed, whilst the criteria of \citet{bai2002determining} are well used, there is still lively debate about what is an appropriate number of factors, and this usually determined by a mix of domain (prior) knowledge and heuristics such as those presented above. The heuristics provided here seem reasonable in the applications and experiments we consider, however, we do not claim they are optimal for all scenarios.
\end{remark}

\section{Implementation}\label{sec:5}

We have implemented the estimation routine as part of the R package \verb|sparseDFM| available via CRAN. The EM routine and ADMM updates are implemented in C++ using the Armadillo library. Initialisation of the ADMM iterates utilises a warm start procedure whereby the solution at the previous iteration of the EM algorithm initialises the next solution. Furthermore, warm-starts are utilised when searching over an $\alpha$ tuning grid. As noted in other applications \citep{Hu2016} starting the ADMM procedure can lead to considerable speed-ups. With regards to the augmentation parameter $\nu$ in the ADMM algorithm, we simply keep this set to $1$ for the experiments run here, however, it is possible that tuning this parameter could lead to further speedups.

On the first iteration of the algorithm, the EM procedure is initialised by a simple application of PCA to the standardised data, analagously to how the preliminary factors and loadings $\bar{\blam}$ were found in Section \ref{sec:tuning}. A summary of the EM algorithm as a whole is given in Algorithm \ref{algo1}.


\begin{algorithm}[h]
\caption{Sparse DFM - EM Algorithm}\label{algo1}
\begin{algorithmic}[1]
\Require $\bX$, $\alpha$
\Ensure $\blam$, $\bA$, $\bsig_{\epsilon}$, $\bsig_{u}$
\State Initialize $\btheta = (\blam, \bA, \bsig_{\beps}, \bsig_{\bu})$ via cubic spline fitting (for missing value imputation) followed by PCA and a VAR fit
\Repeat
\Comment{E-Step}
\State Obtain $\ba_{t\vert n}$ and $\bP_{t\vert n}$ via univariate Kalman filtering and smoothing
\State Calculate $\bP_{t,t-1\vert n}$ via Eq.~\eqref{eq:deJong}
\Comment{M-Step}
\State Update $\bA$ and $\bsig_{\bu}$ via Eqs.~\eqref{eq:estA} and \eqref{sigu}
\State Initialize $\blam^{(0)}=\bZ^{(0)}=\bU^{(0)}=0$
\For{$k=0,...,$ until convergence}
\State $\blam^{(k+1)}=\argmin_{\blam}\mathcal{C}(\blam,\bZ^{(k)},\bU^{(k)})$ via Eqs.~\eqref{eq:Lambda_sparse} and \eqref{eq:fast_inv}
\State $\bZ^{(k+1)}=\mathrm{soft}(\blam^{(k+1)}+\bU^{(k)};\alpha/\nu)$
\State $\bU^{(k+1)}=\bU^{(k)}+\blam^{(k+1)}-\bZ^{(k+1)}$
\EndFor
\State Update $\bsig_{\beps}$ via Eq.~\eqref{eq:sigeps}
\Until{convergence}
\end{algorithmic}
\end{algorithm}

\section{Synthetic Experiments}\label{sec:6}

We provide a Monte-Carlo numerical study to show the performance of our QMLE estimator in terms of recovery of sparse loadings and the ability of the sparse DFM to forecast missing data at the end of the sample. In particular, we simulate from a ground-truth model according to:
\begin{align*}
    \bX_t &= \blam \bF_t + \beps_t \, , \hspace{4em} \beps_t \sim N(\bm{0}, \bsig_{\beps}) \, , \\
    \bF_t &= \bA\bF_{t-1} + \bu_t \, , \hspace{3em}\bu_t \sim N(\bm{0}, \bsig_{\bu}) \, ,
\end{align*}
for $t=1,\dots,n$ and $\bX_t$ having $p$ variables. We set the number of factors to be $r=2$ and consider true model parameters of the form: 
\begin{align*}
    & \blam = \bI_2 \otimes \bm{1}_{p/2} = \begin{bmatrix}
    \bm{1}_{p/2} & \bm{0}_{p/2} \\
    \bm{0}_{p/2} & \bm{1}_{p/2}
    \end{bmatrix} \, , \\
    & \bsig_{\beps} = \bI_p \, , \\
    & \bA = \begin{bmatrix}
    a & 0 \\
    \rho & 0 
    \end{bmatrix} \, , \\
    & \bsig_u = \begin{bmatrix}
    1-a^2 & 0 \\
    0 & 1-\rho^2
    \end{bmatrix} \, .
\end{align*}
The loadings matrix $\blam$ is a block-diagonal matrix which is 1/2 sparse with $p/2$ ones in each block. We set up the VAR(1) process of the factors in this way such that we can adjust the cross-correlation parameter $\rho$ between the factors while having factors that always have variance one. This allows us to understand how important a cross-correlation at non-zero lags structure is when assessing model performance. We vary the $\rho$ parameter between $\rho=\{0,0.6,0.9\}$, going from no cross-correlation to strong cross-correlation between the factors.  We set the covariance of the idiosyncratic errors to be $\bI_p$ in order to have a signal-to-noise ratio between the common component $\blam \bF_t$ and the errors $\beps_t$ equal to one. 

\begin{figure}[h!]
\centering
    \includegraphics[width=0.8\linewidth]{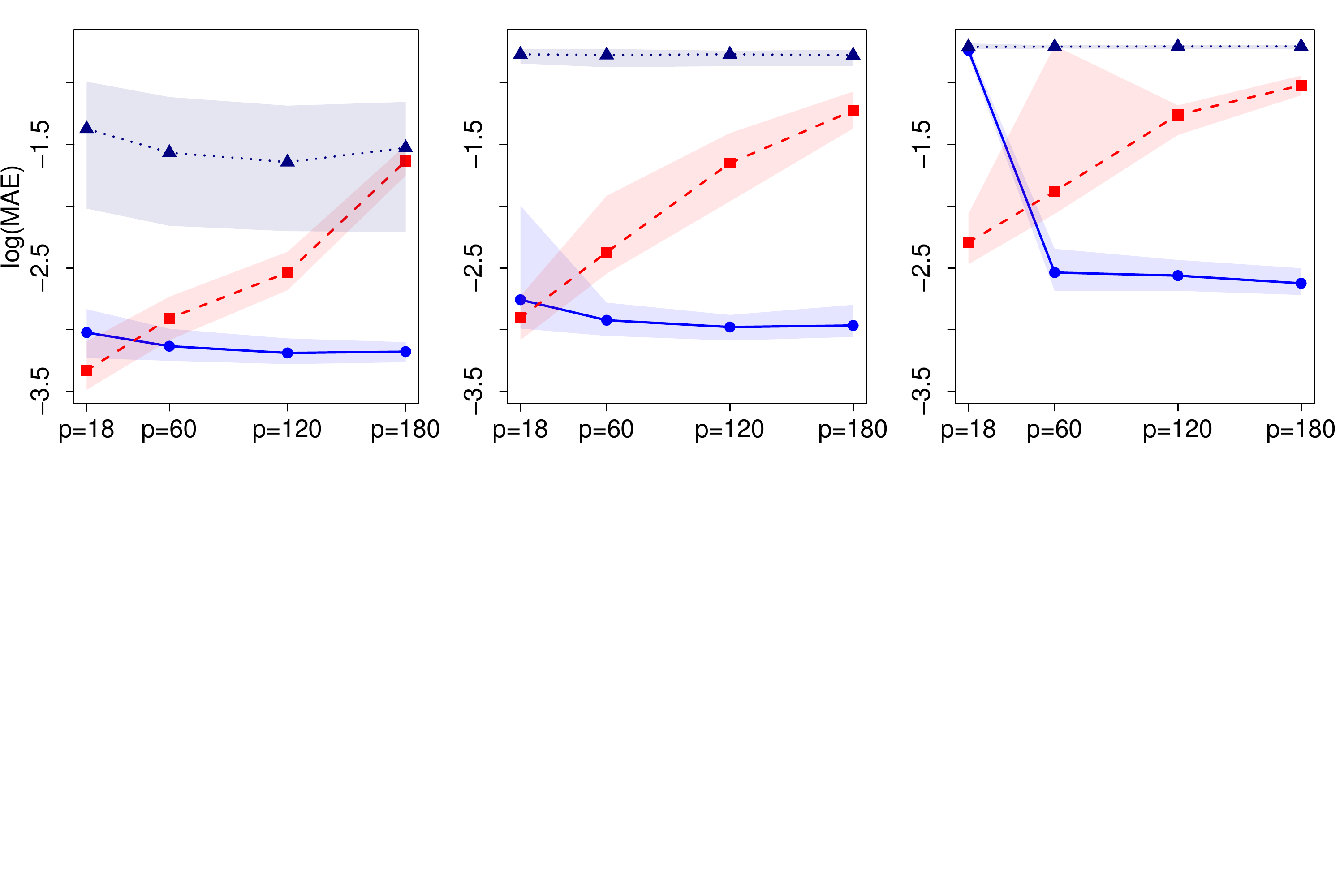} \\
    \vspace{-12em}
    \includegraphics[width=0.8\linewidth]{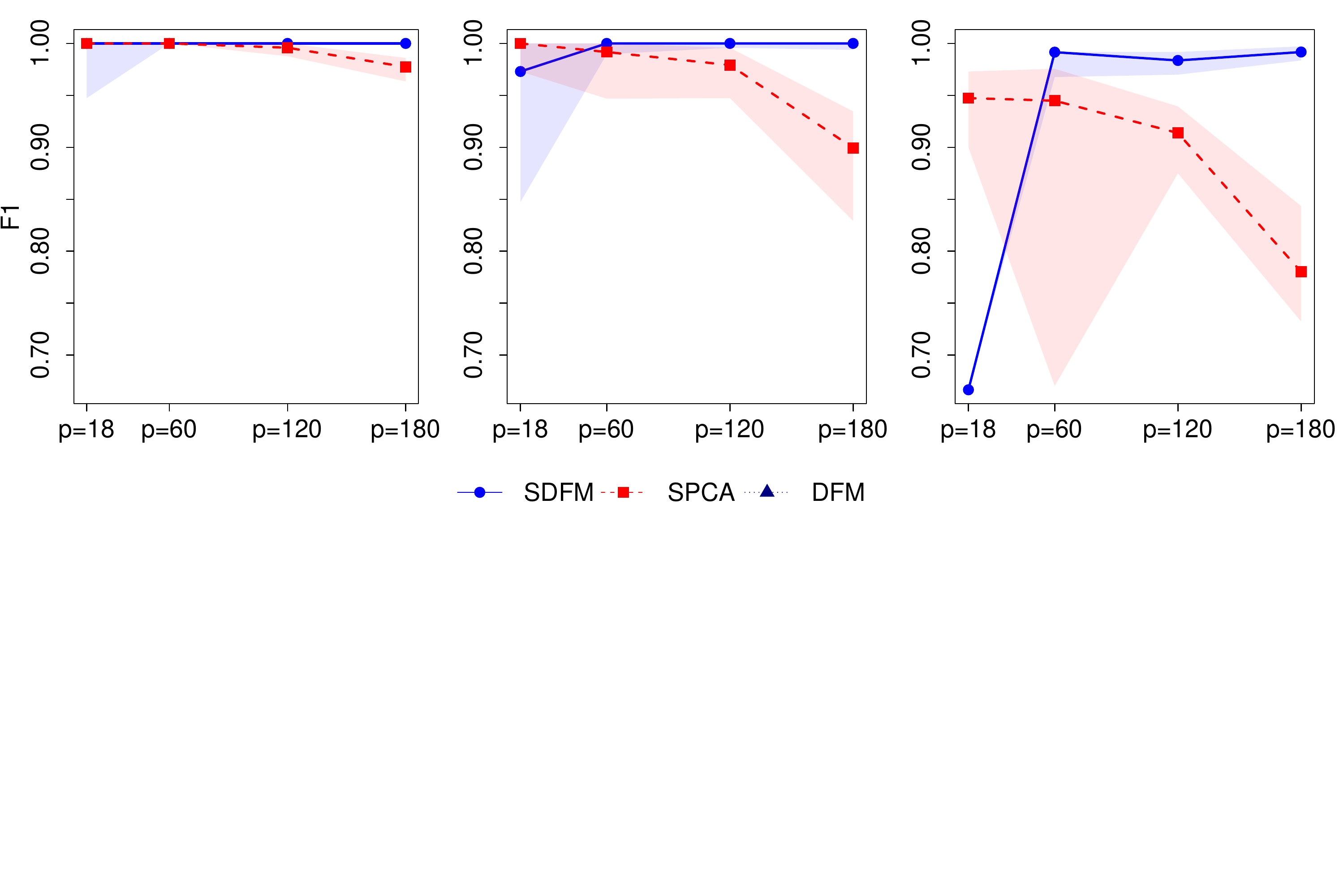}
    \vspace{-11em}
    \caption{Median log-MAE score (top panel) and median F1 score (bottom panel) for recovering factor loadings across 100 experiments with a shaded confidence band of the 25th and 75th percentile. The plots represent a setting with a fixed $n = 100$ and varying number of variables $p$ and where the cross-correlation parameter in the VAR(1) process is set to $\rho = 0$ (left plot), $\rho = 0.6$ (middle plot) and $\rho = 0.9$ (right plot).}
    \label{fig:loadings_recovery}
\end{figure}

\subsection{Recovery of Sparse Loadings}

We apply our sparse DFM (SDFM) to simulated data from the data generating process above to assess how well we can recover the true loadings matrix $\blam$. We compare our method to sparse principle components analysis\footnote{The SPCA algorithm is implemented using the \emph{elasticnet} R package available on CRAN.} (SPCA) applied to $\bX_t$ to test which settings we are performing better in. We tune for the best $\ell_1$-norm parameter in both SDFM and SPCA using the BIC function (\ref{eq:BIC}) by searching over a large grid of logspaced values from $10^{-3}$ to $10^2$. We also make comparisons to the regular DFM approach of \cite{banbura2014maximum} to test the importance of using regularisation when the true loading structure is sparse. 

The estimation accuracy is assessed with mean absolute error (MAE) between the true loadings according to $(rp)^{-1}\|\hat{\blam}-\blam\|_1$. We also provide results for the F1 score for the sparsity inducing methods of SDFM and SPCA to measure how well the methods capture the true sparse structure. Due to invariance issues discussed, the estimated loadings may not be on the same scale as the true loadings, we thus first re-scale the estimated loadings such that their norm is equal to that of the simulated loadings, i.e.~$\|\hat{\blam}\|_2 = \|\blam \|_2$. 
The estimated loadings from each model are identified up to column permutations and therefore we permute the columns of $\hat{\blam}$ to match the true order of $\blam$. We do this by measuring the 2-norm distance between the columns of $\hat{\blam}$ and $\blam$ and iteratively swapping to match the smallest distances. 

Figure \ref{fig:loadings_recovery} displays the results for the loadings recovery where we have fixed the number of observations to be $n=100$ and vary the number of variables between $p=\{ 18, 60, 120, 180\}$ along the x-axis and the cross-correlation parameter in the VAR(1) process between $\rho=\{0,0.6,0.9\}$ going from the left to middle to right plot respectively. The top panel shows the median MAE score (in logarithms) over 100  experiments while the bottom panel shows the F1 scores. We provide confidence bands for both representing the 25th and 75th percentiles. It is clear from the plots that the sparsity inducing methods of SDFM and SPCA are dominating a regular DFM when the true loadings structure is in fact sparse. It is also clear that SPCA performs poorly, compared with SDFM, when the cross-section of the data increases for a fixed $n$. This is even more noticeable from the F1 score when $\rho$ increases. This highlights the importance of the SDFM's ability to capture correlations between factors at non-zero lags. Unlike SPCA, the EM algorithm of SDFM allows feedback from the estimated factors when updating model parameters, allowing it to capture these factor dependencies. We see improved scores in MAE as the cross-section increases for SDFM. This follows the intuition of the EM algorithm framework as we learn more about the factors as the dimension $p \rightarrow \infty$. We should remark that for most scenarios the F1 score for SDFM is almost one, however, when $p=18$ and $\rho$ is high, the score does drop. In this setting a low value for $\alpha$ minimises BIC, meaning almost no sparsity is applied (a very similar result to a regular DFM fit). Here, the two factors are highly correlated and there is not enough cross-section to determine factor structure. In practice it is likely that cross-section will be large and hence this result is not too concerning.

\begin{figure}[h]
\centering
    \includegraphics[width=0.8\linewidth]{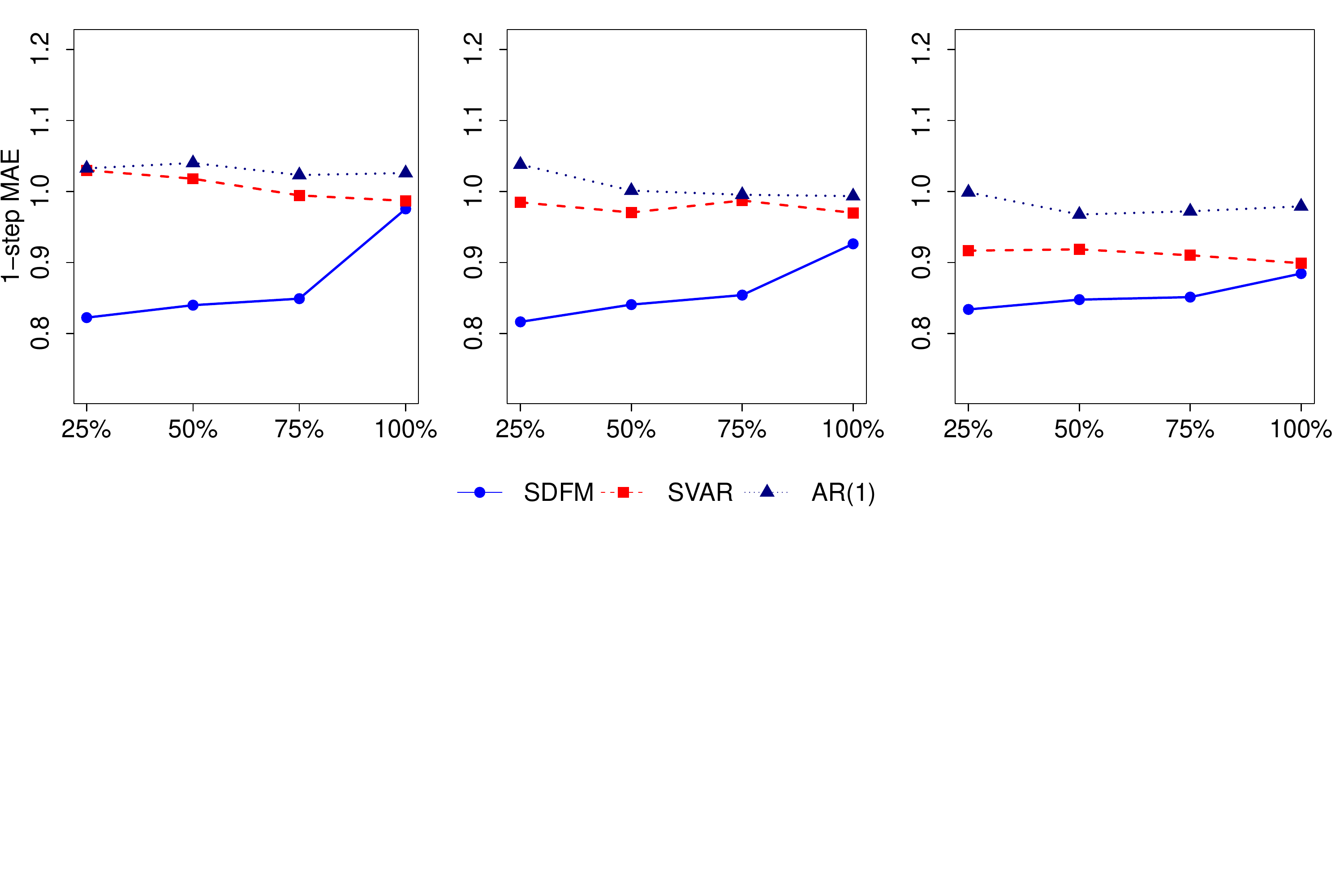} \\
    \vspace{-10em}
    \caption{Average MAE score forecasting, as a function of the level of missing data in the last sample. From left-right: $\rho=0$, $\rho=0.6$, $\rho=0.9$. Plot indicates the 50th percentile of performance across 100 experiments with $n=100$, $p=64$.}
    \label{fig:forecast_accuracy}
\end{figure}

\subsection{Forecasting Performance}
To evaluate our ability to forecast missing data at the end of the sample we simulate data from the data generating process above with $n = 200$, $p=64$ and consider $\rho=\{0,0.6,0.9\}$, and assume different patterns of missing data at the end of the sample. We consider a 1-step ahead forecast case where we set $25\%$, $50\%$, $75\%$ and then $100\%$ of variables to be missing in the final row of $\bX$. When allocating variables to be missing we split the data up into the two loading blocks and set the first $25\%$, $50\%$, $75\%$ and $100\%$ of each loading block to be missing. For example, the variables 1 to 8 and 33 to 40 are missing in the $25\%$ missing scenario. We are interested in forecasting the missing data in the final row of $\bX$ and we calculate the average MAE over 100 experiments. 

We make comparisons with a sparse vector-autoregression (SVAR) model\footnote{The SVAR algorithm is implemented using the \emph{BigVAR} R package available on CRAN. This has a built-in cross-validation mechanism to tune for the best $\ell_1$-penalty parameter which we use in our simulations.} as this is a very popular alternative forecasting strategy for high-dimensional time series that is based on sparse assumptions. As our factors are generated using a VAR(1) process with a sparse auto-regression matrix, we are interested to see whether SVAR will be able to capture the cross-factor auto-correlation when producing forecasts. We also apply a standard AR(1) process to each of the variables needing to be forecasted as a benchmark comparison. 

Figure \ref{fig:forecast_accuracy} displays the results of the simulations plotting MAE for each of the 3 methods and each simulation setting. In all settings we find SDFM to outperform both SVAR and AR(1). When $\rho$ is set to be $0.9$, we find SVAR does improve its forecasting performance as opposed to when $\rho=0$ as the VAR(1) process driving the factors becomes more prominent. The results confirm SDFM's ability to make use of variables that are present at the end of the sample when forecasting the missing variables. We see this by the rise in MAE when $100\%$ of the variables are missing at the end of the sample and the model can no longer utilise available data in this final row. The MAE remains fairly flat as the amount of missingness rises from $25\%$ to $75\%$ showing SDFM's ability to forecast correctly even when there is small amount of data available at the end of the sample. 

\subsection{Computational Efficiency}

\begin{figure}[h]
\center
\includegraphics[width=0.35\linewidth]{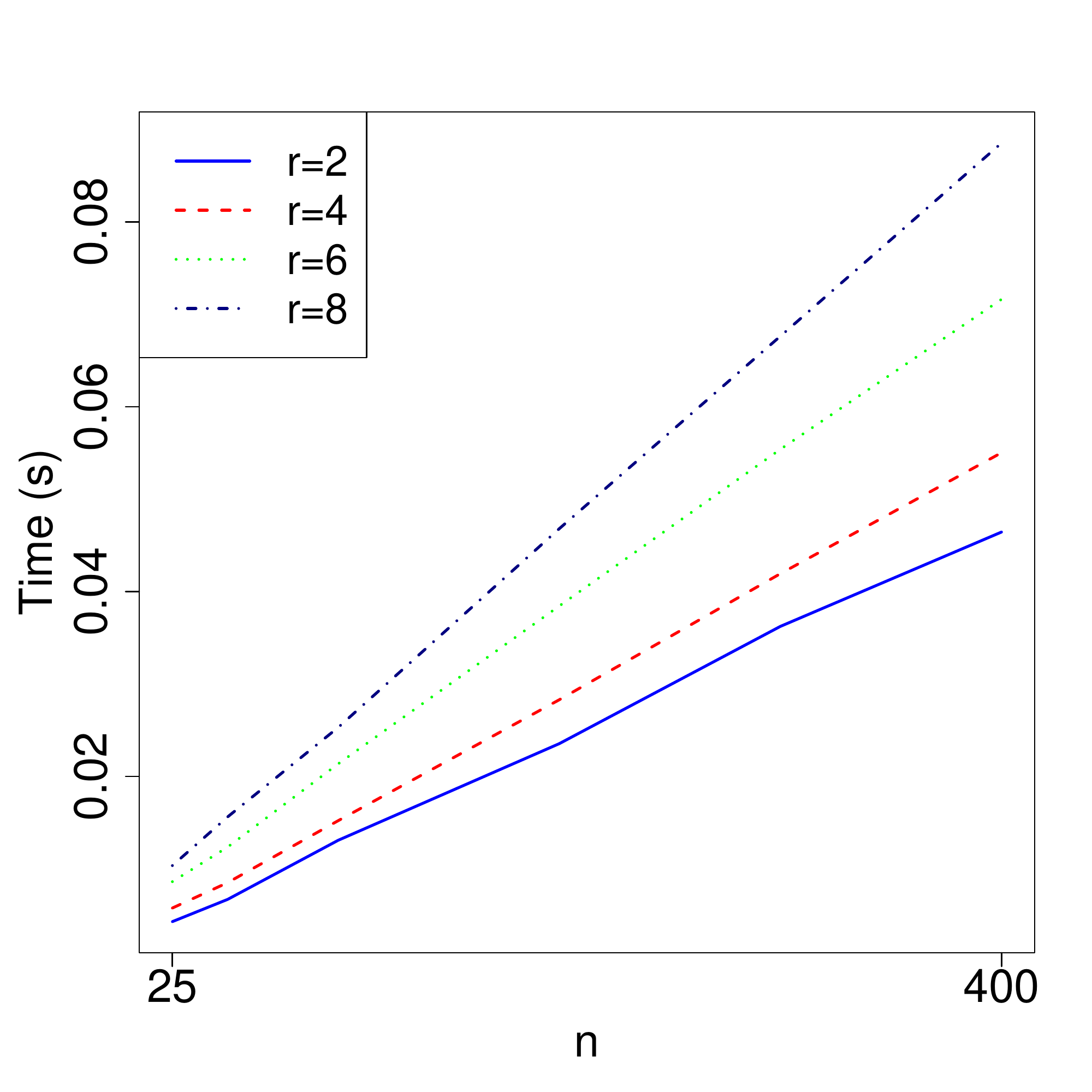}
\includegraphics[width=0.35\linewidth]{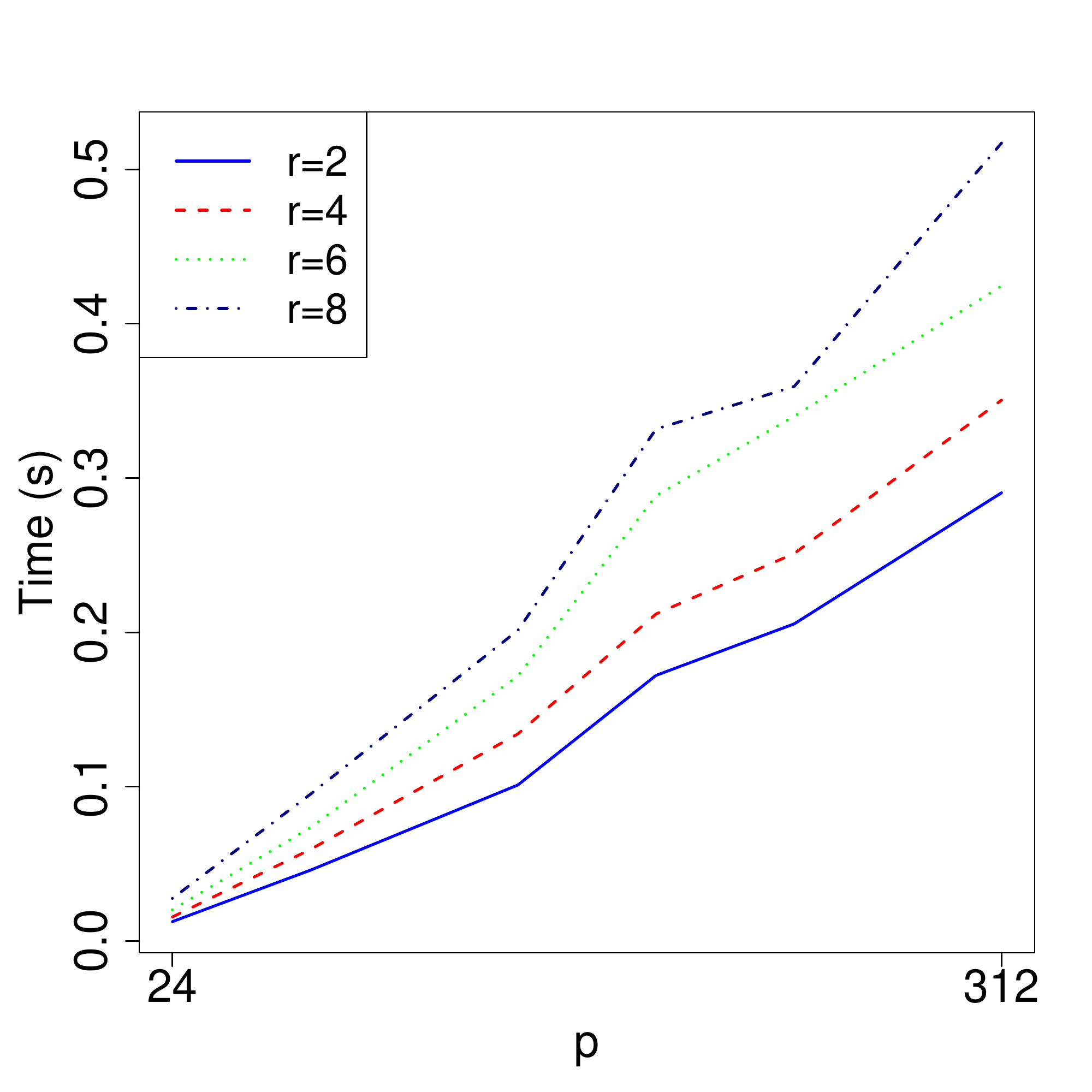}
\caption{Summary of computational cost. Top: as a function of $n$, with fixed $p=24$. Bottom: as a function of $p$, with fixed $n=100$. Average performance across 10 experiments.}
\label{fig:timings}
\end{figure}

%

To assess the computational scalaibility, we simulate from a sparse DFM where $\blam = \bI_r \otimes \bm{1}_{p/r}$ and $\bsig_\epsilon = \bI_p$, and the factors are a VAR(1) with $\bA=0.8 \times \bI_r$ and $\bsig_u = (1-0.8^2) \times \bI_r$. We record the number of EM iterations and the time they take for each $\ell_1$-norm parameter $\alpha$ up to the optimal $\ell_1$-norm parameter $\hat{\alpha}$ and then take the average time of a single EM iteration. We repeat the experiment ten times for each experimental configuration. 

The results are presented in Figure \ref{fig:timings}, which demonstrates scalability as a function of $n$, and $p$, under different assumptions on the number of factors $r=2,4,6,8$. 
As expected, the cost is approximately linear in $n$ and $p$, with increasing cost as a function of the number of factors $r$. The results demonstrate the utility of using the univariate smoothing approach as well as the matrix decomposition when calculating required inversions.

\begin{figure*}[h]
\includegraphics[width=\linewidth]{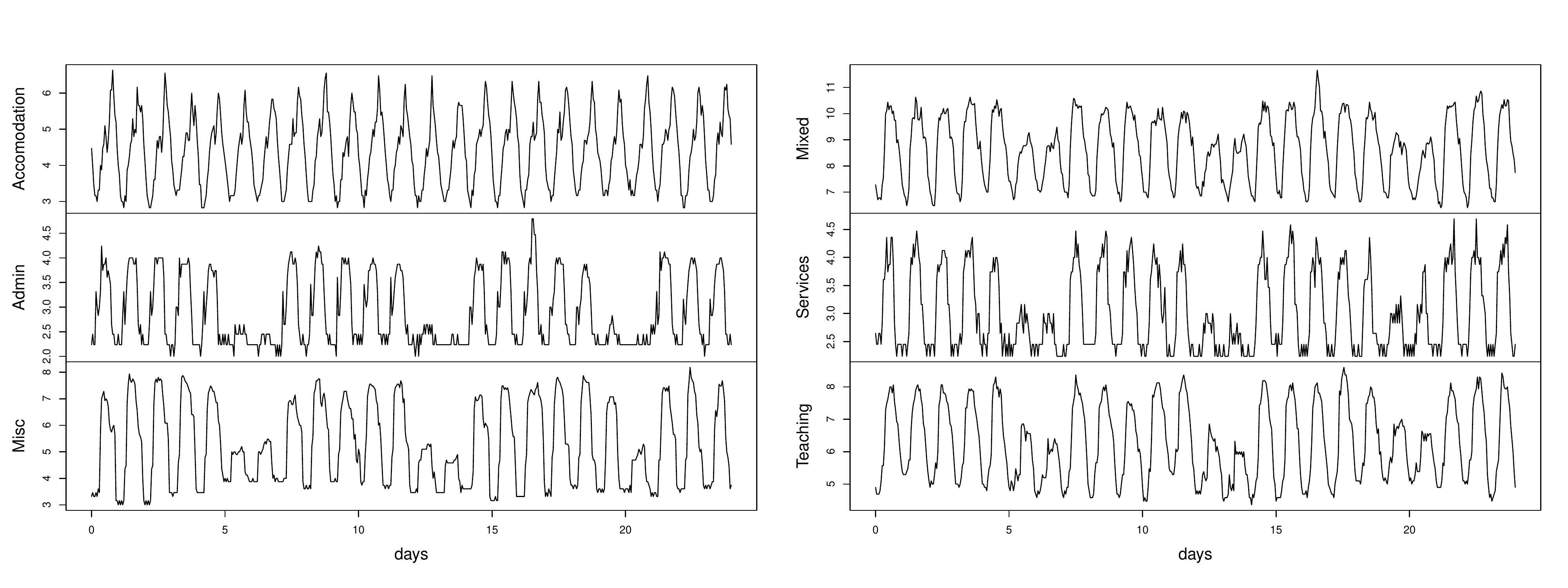}
\caption{Example of time-series readings for the 24 days under analysis. The figures present the square root of the consumption in each hour ($\sqrt{6\mathrm{kWh}}$) for different types of building, and illustrate the diverse nature of consumption.}\label{fig:examples}
\end{figure*}

\section{The Dynamic Factors of Energy Consumption}\label{sec:7}

This section details application of the sparse DFM to a real-world problem, namely the forecasting and interpretation of energy consumption. Beyond forecasting consumption in the near-term future, our aim here is to also characterise the usage in terms of what may be considered typical consumption profiles. These are of specific interest to energy managers and practitioners, as understanding how energy is consumed in distinct buildings can help target interventions and strategy to reduce waste. We also highlight how the sparse DFM, and in particular our EM algorithm, can be used to impute missing data and provide further insight.

\subsection{Data and Preprocessing}

In this application, the data consists of one month of electricity consumption data measured across $p=42$ different buildings on our universities campus. This data is constructed based on a larger dataset, which monitors energy at different points throughout a building, in our case, we choose to aggregate the consumption so that one data-stream represents the consumption of a single building. The data is gathered at 10 minute intervals (measuring consumption in kWh over that interval), resulting in $n=3,456$ data points spanning 24 days worth of consumption in November 2021, we further hold out one day $n_{\mathrm{test}}=144$ data points to evaluate the out-of-sample performance of the DFM model. An example of time series from the dataset is presented in Figure \ref{fig:examples}. There are many alternative ways one may wish to model this data, however, one of the key tasks for energy managers is to understand how consumption in this diverse environment is \emph{typically} structured.  This is our primary objective in this study, i.e.~we wish to extract typical patterns of consumption that can well represent how energy is used across the campus. To this end, we decide not to remove the relatively clear seasonal (daily) patterns in consumption prior to fitting the factor model, the hope being, that these patterns will somehow be pervasive in the derived factors.


Whilst we do have meta-data associated with each of these buildings for sensitivity purposes we choose to omit this in our discussions here, the buildings are presented as being approximately categorised under the following headings:

\begin{description}
\item[Accomodation:] Student residences, and buildings primarily concerned with accomodation/student living.
\item[Admin:] Office buildings, e.g.~HR, administration, and central university activities.
\item[Misc:] Other student services, e.g.~cinema, shopping, sports facilities.
\item[Mixed:] Buildings which mix teaching and accomodation. For instance, seminar rooms on one floor with accomodation on another.
\item[Services:] Management buildings, porter/security offices.
\item[Teaching:] Teaching spaces like lecture theatres, seminar rooms.
\end{description}


\subsection{Factor Estimates and Interpretation}

\begin{figure}[h]
\centering\includegraphics[width=0.5\linewidth]{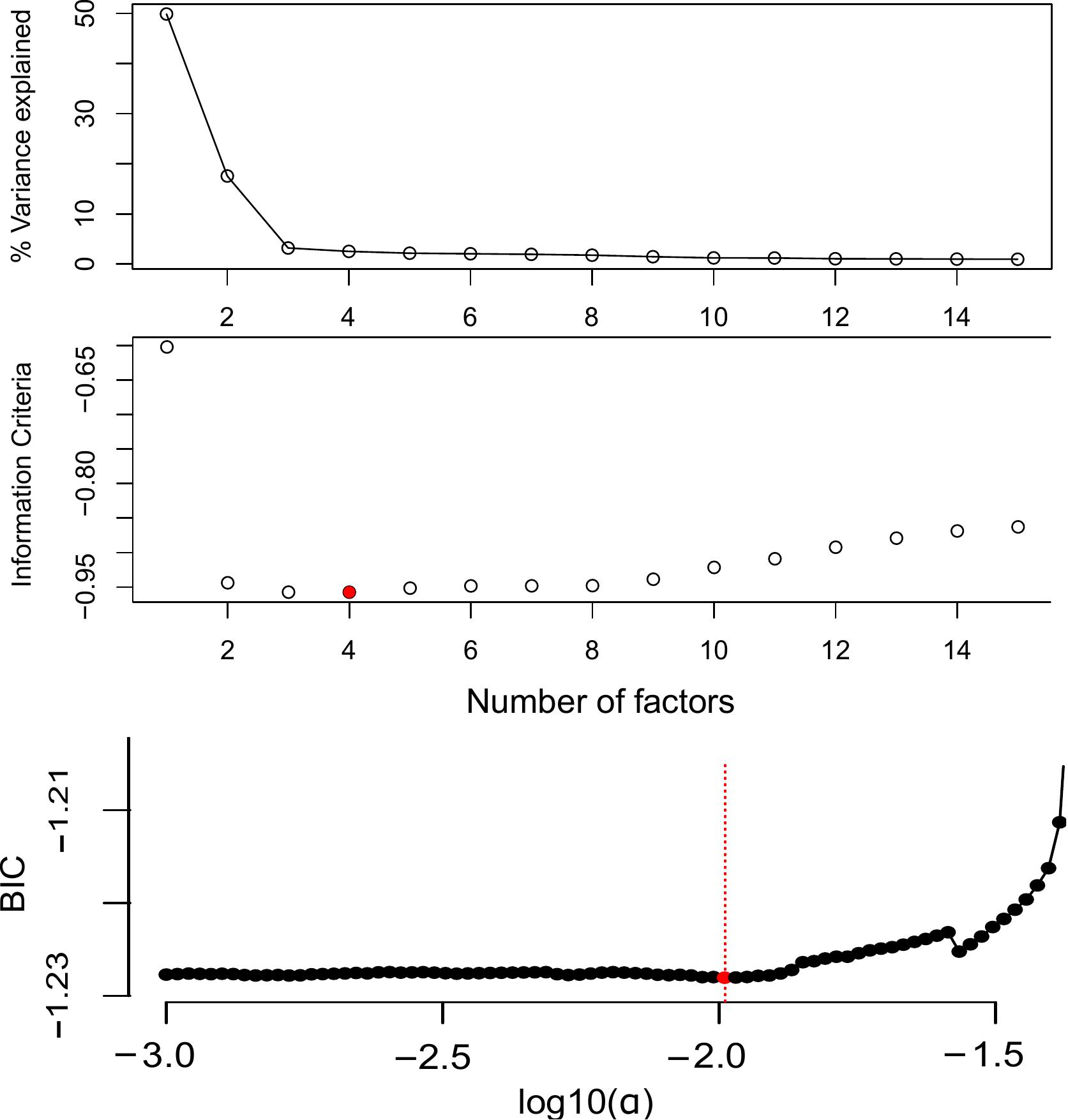}
\caption{Top: Proportion of variance explained, based on PCA applied to the scaled and pre-processed (interpolated) dataset. Middle: Information Criteria (\ref{eq:IC2}) as a function of number of retained factors $r$. Bottom: BIC as a function of $\alpha$, vertical line indicates minimiser and the $\alpha$ used in the subsequent analysis.}
\label{fig:ic_combined}
\end{figure}

\begin{figure}[h]
\centering\includegraphics[width=\linewidth]{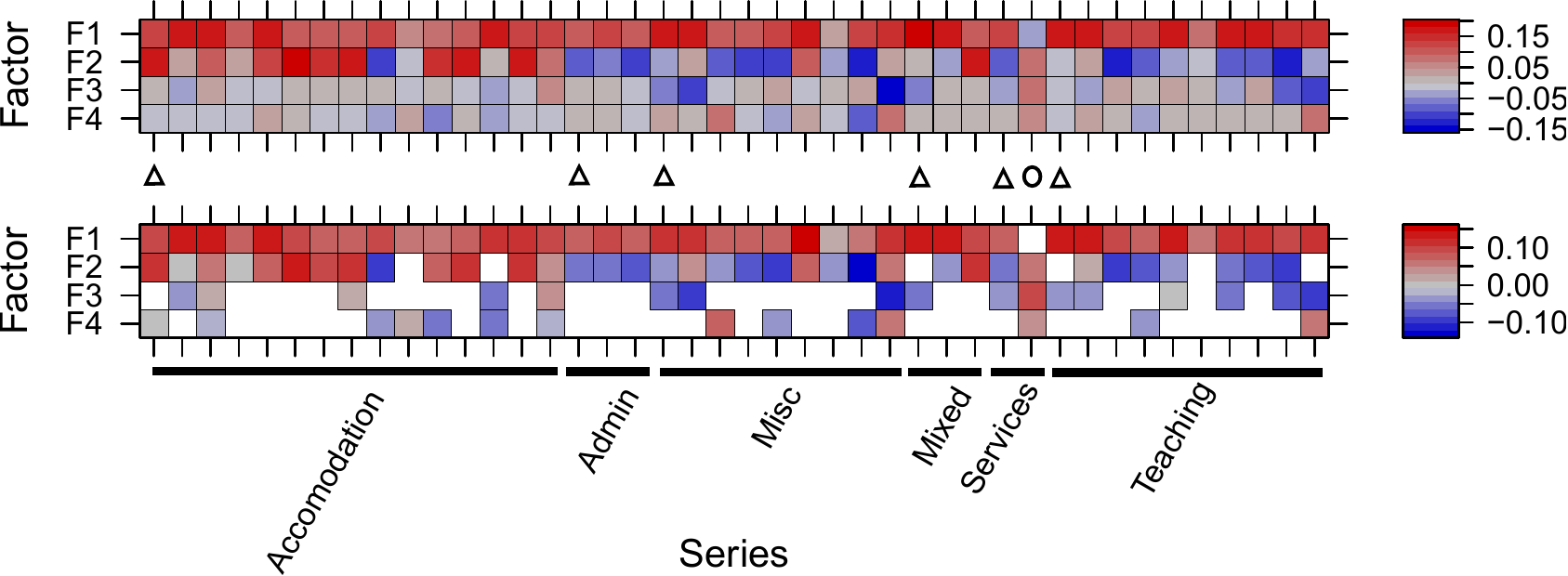}
\caption{Estimated factor loadings for the regular DFM (top) and sparse DFM (bottom). Series are categorised according to one of six building types, triangles indicate the example series plotted in Fig \ref{fig:examples}.}\label{fig:heatmap_energy}
\end{figure}

\begin{figure}[h]
\centering\includegraphics[width=1\linewidth]{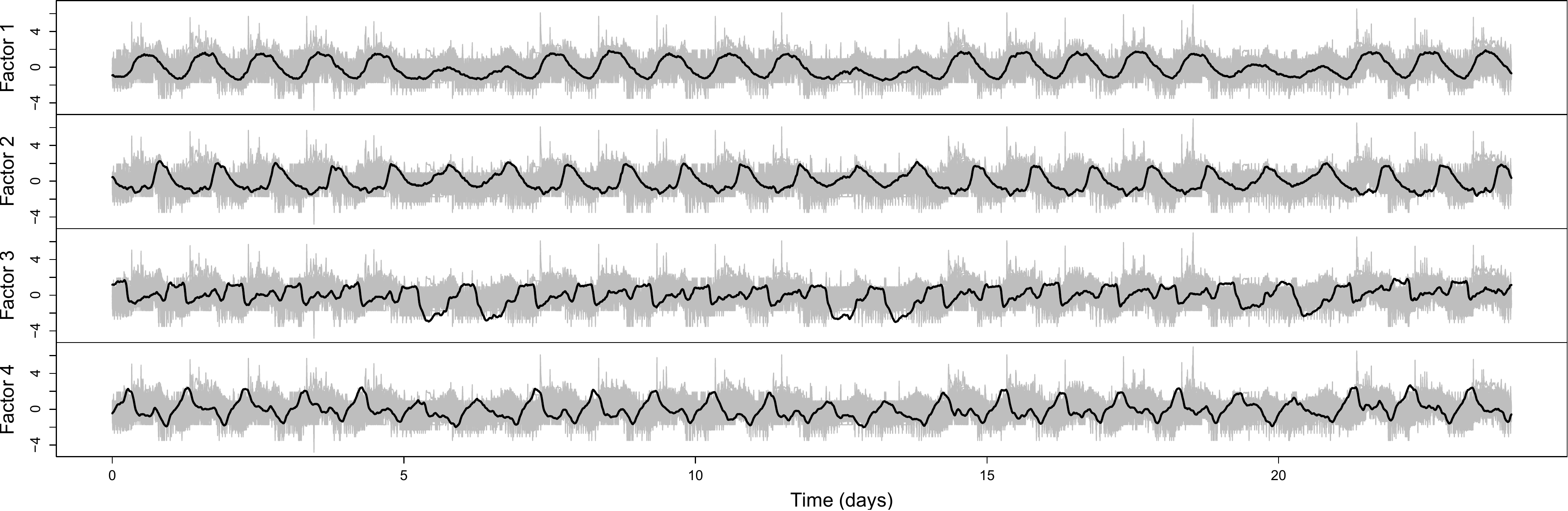}
\caption{Estimated Factors (black) with original data (grey) as a function of time using the optimal $\alpha=0.01$ chosen according to BIC. When multiplied by the factor loadings (top) gives the estimated common component.}\label{fig:factor_energy}
\end{figure}

To estimate factors we first choose a number of factors according to criterion (\ref{eq:IC2}), which leads to 4 factors being specified as seen in Fig.~\ref{fig:ic_combined}. Next, we apply the sparse DFM model via the EM procedure in Algorithm \ref{algo1}. We run the algorithm to scan across a range of $\alpha$ parameters, and in this case, the BIC criteria suggests to impose moderate sparsity corresponding to $\alpha\approx 0.01$. One may note in Figure \ref{fig:ic_combined} that there is a second dip in the BIC criteria around $\alpha\approx0.03$ after which the BIC rapidly rises until the cutoff constraint, after which all $\hat{\Lambda}_{ij}$ are set to zero. In this case, the sparsity pattern of the two above values of $\alpha$ appear very similar, and the loading of the variables on the factors appears largely stable as a function of $\alpha$. To give some intuition, the loadings $\hat{\blam}$ for $\alpha=0.01$ and $\alpha=0$ are visualised in Fig \ref{fig:heatmap_energy}, a visualisation for the corresponding factors $\ba_{t\vert n}$ are given in Fig.~\ref{fig:factor_energy}.

\begin{figure}[h]
\centering
\includegraphics[width=0.6\linewidth]{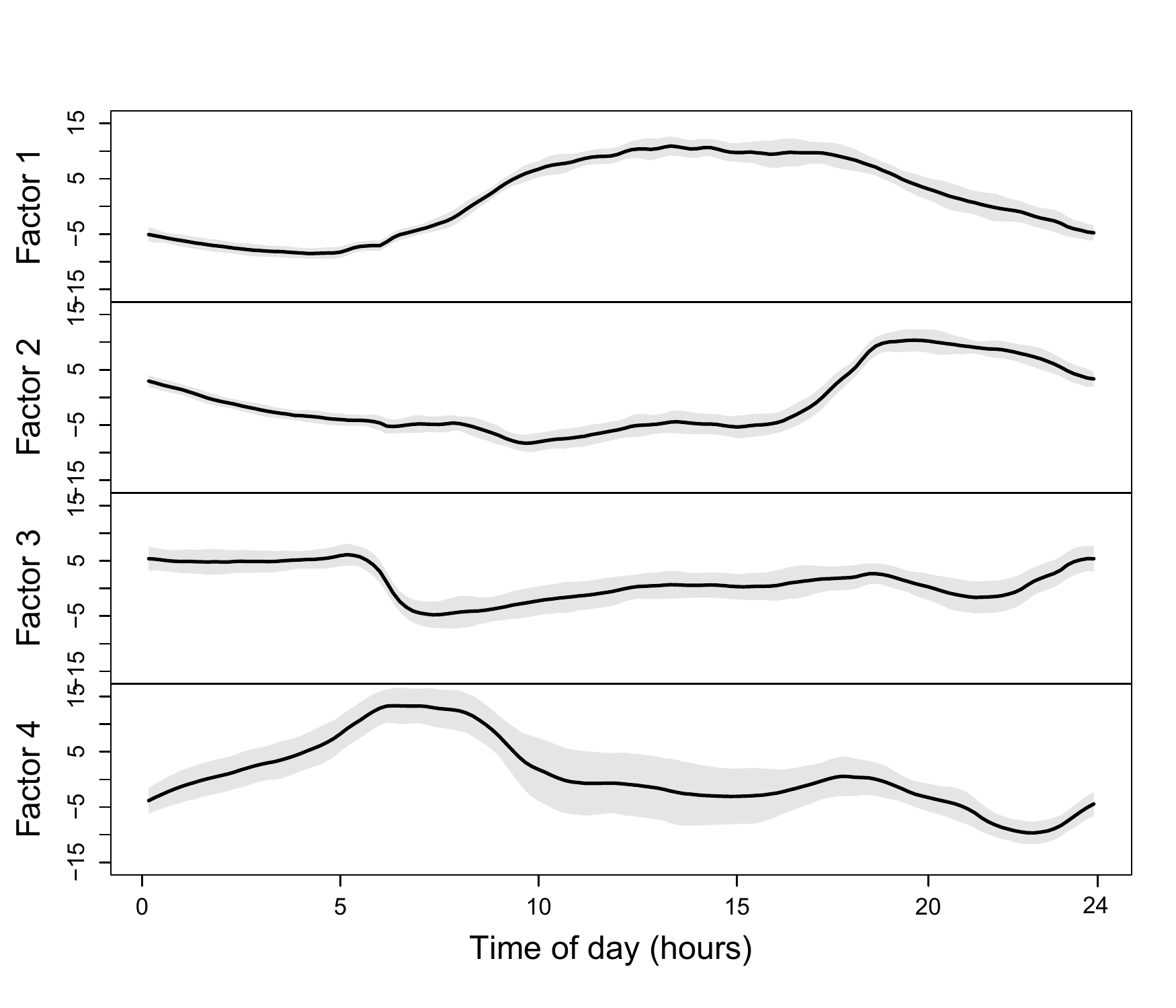}
\caption{Average factor profile as a function of time-of-day, $t=0$ corresponding to midnight. The solid line is a pointwise average of the factor $\hat{\ba}_{t\vert n}$ across the 18 weekdays in the sample, confidence intervals are constructed as $\pm 1.96$ the standard-deviation.}
\label{fig:day_average}
\end{figure}

For brevity, we focus on analysing the results of the spare DFM model. Of particular interest for the energy manager is the interpretation of consumption that the sparse DFM model provides, and is most obvious for the third and fourth factors in this case. A visualisation of the factor behaviour on a typical weekday is given in Figure \ref{fig:day_average} where there is a clear ordering in the uncertainty surrounding the factor behaviour, e.g.~Factor one has small confidence intervals, whereas Factor 4 has more uncertain behaviour, especially during the working day. Interestingly, the sparse DFM only really differs from the regular DFM in these third and fourth factors, where the latter exhibits slightly greater variation in behaviour. The sparse DFM is able to isolate these further factors to specific buildings. For example, the building identified by the circle in Fig.~\ref{fig:heatmap_energy} is known to be active primarily throughout the night, and we see its factor loadings reflect this, e.g.~the regular working day cycles for Factor 1 are not present, however, the evening and early morning features (Factors 3, and 4) are represented. For the teaching buildings, we see that the loading on Factor 2, and 3, are negative, indicating a sharp drop-off in energy consumption in the evening/overnight, again, this aligns with our expectations based on the usage of the facilities.

\subsection{Forecasting Performance}

\begin{figure}
\includegraphics[width=\linewidth]{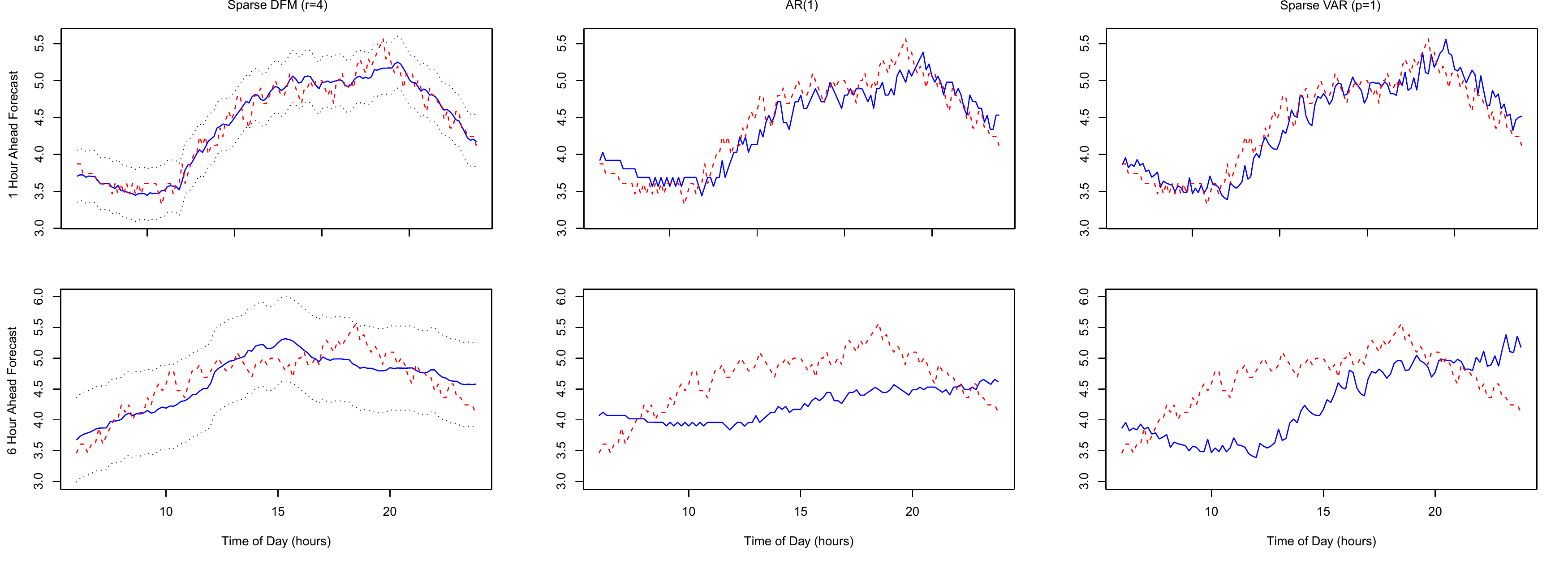}
\caption{Example of predicted consumption ($\sqrt{\mathrm{kWh}}$) in one (accommodation) building on the campus. The top row represents 1 hour ahead forecasts based on an expanding window, whilst the bottom represents 6 hour ahead forecasts. The SDFM and SVAR are tuned on the 24 days of data prior to that presented in the figure. Confidence intervals for the SDFM are based on $1.96\times[\hat{\blam}\bP_{t\vert n} \hat{\blam}^\top + \hat{\bsig}_{\epsilon}]_{ii}^{1/2}$}.
\label{fig:example_forecast}
\end{figure}

\begin{figure}
\includegraphics[width=1\linewidth]{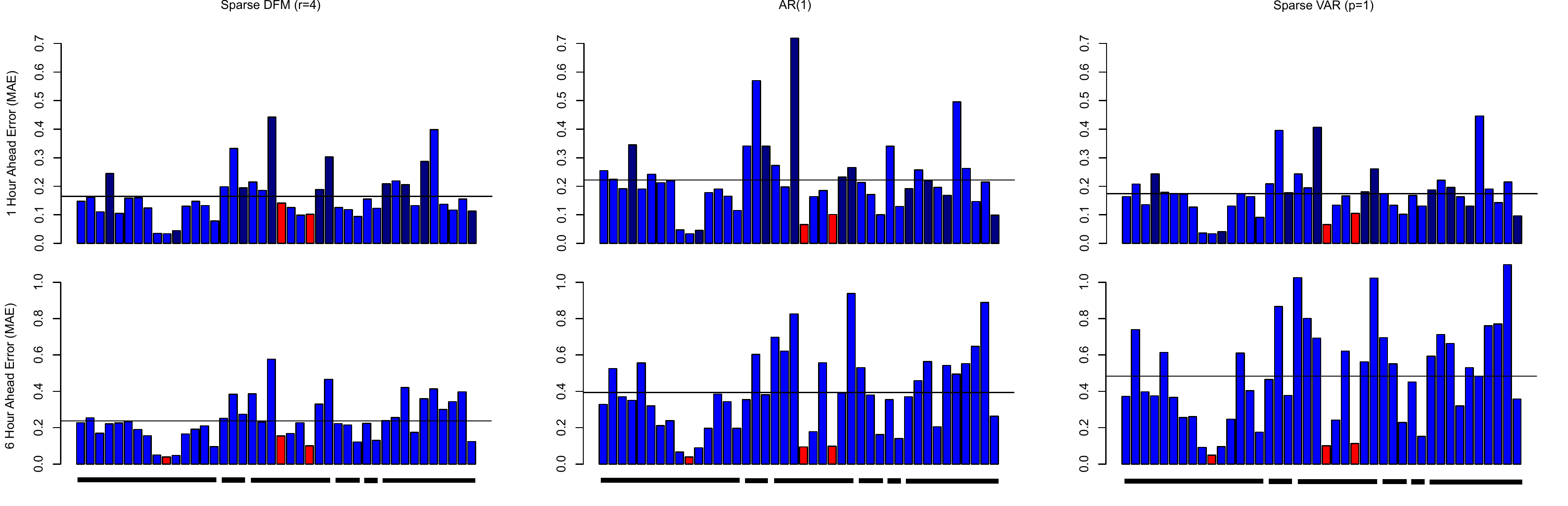}

\caption{Forecast Errors (MAE) for each building for (top) 1 hour ahead forecast, and (bottom) 6 hour ahead forecast. Performance evaluated on one hold out day $144-h$ data points. Each bar is colored according to which method performs best for that building. Blue: SDFM, Red: AR(1), Navy: SVAR. The solid black line indicates average performance across all buildings, the grouping of buildings is indicated via the dashed line under the plots.}.
\label{fig:forecast_eval}

\end{figure}

The primary motivation for applying the sparse DFM in the context of this application is to aid in interpreting the consumption across campus. However, it is still of interest to examine how forecasts from the DFM compare with competitor methods. For consistency, we here provide comparison to the AR(1) and sparse VAR methods detailed earlier. These models all harness a simple autoregressive structure to model temporal dependence, specifically regressing only onto the last set of observations (or factors), i.e.~they are Markov order 1. Our experiments asses performance of the models in forecasting out-of-sample data, either $h=6$ steps ahead (1 hour), or $h=36$ steps ahead (6 hours). The forecasts are updated in an expanding window manner, whereby the model parameters are estimated on the 24 days of data discussed previously, the forecasts are then generated sequentially based on $n+t=1,\ldots,n_{\mathrm{test}}=144-h$ observations. An example of the forecasts generated (and compared to the realised consumption) is given in Figure \ref{fig:example_forecast}. A striking feature of the DFM based model is its ability to (approximately) time the increases/decreases in consumption associated with the daily cycle. These features in the AR(1) and sparse VAR model are only highlighted after a period of $h$ steps has passed, e.g.~the models cannot anticipate the increase in consumption.

A more systematic evaluation of the forecast performance is presented in Figure \ref{fig:forecast_eval}, where the average error is calculated for each building, for each of the different models. We see that for the 1 hour ahead forecasts, all methods perform similarly, with the sparse DFM winning marginally, and the AR(1) forecasts demonstrating more heterogeneity in the performance. There is no clear winner across all the buildings, for most (30) buildings the DFM forecasts prove most accurate, with the AR being best on 2, and the SVAR winning on the remaining 10. Moving to the 6 hour ahead forecasts, the dominance of the sparse DFM becomes clear, winning across 39 of the buildings, and the AR method winning on 3. Interestingly, the SVAR fails to win on any building, falling behind the simpler AR approach. This suggests, that in this application the activity of one building may not impact that of another across longer time-frames, however, the behaviour of the latent factors (common component) does provide predictive power. 

One could reasonably argue that we should not use these competitor models in this way for forecasting, e.g. we would likely look to add seasonal components corresponding to previous days/times, and/or potentially a deterministic (periodic) trend model. However, these extensions can also potentially be added to the DFM construction. Instead of absolutely providing the best forecasts possible, this case-study aims instead to highlight the differences in behaviour across the different classes of models (univariate, multivariate sparse VAR, and sparse DFM), and the fact that the sparse DFM can borrow information from across the series in a meaningful way, not only to aide interpretation of the consumption, but also to provide more accurate forecasts by harnessing the common component.

\section{Conclusion}\label{sec:8}

In this paper, we have presented a novel method for performing inference in sparse Dynamic Factor models via a regularised Expectation Maximisation algorithm. Our analysis of the related QML estimator provides support for its ability to recover structure in the factor loadings, up to permutation of columns, and scaling. To our knowledge this is the first time the QMLE approach has been studied for the sparse DFM model, and our analysis extends recent investigations by \cite{despois2022identifying} using more simplistic sparse PCA based approaches. When factors are thought to be dependent, e.g. as in our VAR(1) construction, the QMLE approach appears particularly beneficial relative to SPCA. We also validate that simple BIC based hyper-parameter tuning strategies appear to be able to provide reasonable calibration of sparsity in the high-dimensional setting.

There is much further work that can be considered for the class of sparse DFM models proposed here, for example looking at developing theoretical arguments on consistency, of both factor loadings, and the factor estimates themselves. In this paper, we opted for an empirical analysis of the EM algorithm, which we believe is more immediately useful for practitioners. Theoretical analysis of the proposed estimation routine is challenging for several reasons. First, one would need to decide whether to analyse the theoretical minimiser (QMLE), or the feasible estimate provided by the EM algorithm. Second, we need to consider the performance as a function of both $n$ and $p$. For example, Proposition 2 from \citet{barigozzi2022quasi} gives theoretical results for the consistency of factor loadings for the regular unregularised QMLE and for a dense DFM model. A further line of work would be to generalise these results to the sparse DFM setting, for instance, can we show a result analogous to Theorem 1 in \citet{Bai2016}, that shows the QMLE estimator of the loadings is equivalent to the OLS estimator applied to the true factors? These kind of approaches could potentially enable a formal comparison of the sparse PCA based approaches and our QMLE approach.

On a more methodological front, one could consider extending the regularisation strategy presented here to look at different types of sparsity assumption, or indeed to encode other forms of prior. Two potential extensions could be to relax the assumption that the factor loadings remain constant over time, or adopt a group-lasso type regularisation on the loadings. The latter would enable users to associate factors with pre-defined sub-sets of the observerd series, but still in a somewhat data-driven manner. For instance, in the energy application we could consider grouping the series via type of building and encouraging sparsity at this grouped level, rather than at the building level. This could be particularly useful if we consider the application to smart-meters at the sub-building, e.g. floor-by-floor, or room-by-room level. One of the benefits of the ADMM optimisation routine developed here is that it easily extended to these settings.

A final contribution of our work is to demonstrate the application of the sparse DFM on a real-world dataset, namely the interpretation and prediction of smart meter data. Traditionally, application of DFM based models has been within the economic statistics community, however, there is no reason they should not find much broader utility. The application to modelling energy consumption in a heterogeneous environment is novel in itself, and serves to raise awareness of how the DFM can help provide an exploratory tool for complex high-dimensional time series. In this case, not only is the sparse DFM beneficial for interpreting consumption patterns, identifying distinctive profiles of buildings that qualitatively align with our intuition, e.g. based on type of use, but also in forecasting consumption ahead of time. With the latter, the DFM can borrow from buildings with similar consumption profiles to better predict consumption peaks/dips further ahead in time.

Finally, we would like to remark that further applications of our proposed sparse DFM estimator can be found in our paper \citep{mosleyJSS}, that also provides guidance on how to implement the methods in R. In particular, the application of the DFM to predicting trade-in-goods flows demonstrates that assuming sparse factors can improve forecast performance relative to the DFM, and  that the structure of the loadings can be substantially altered as a function of $\alpha$.

\subsection*{Supplementary Information}
%
Code to replicate the smart-meter example presented in this paper can be found on GitHub (\href{github.com/alexgibberd}{https://github.com/alexgibberd}). The sparse DFM package used to implement the EM algorithm can be found on CRAN, or via Github (\href{github.com/mosleyl/sparseDFM}{https://github.com/mosleyl/sparseDFM}). We refer the reader to \citet{mosleyJSS} for further details on how to use the package.

\subsection*{Acknowledgments}
A.~Gibberd and T.-S.~T.~Chan acknowledge funding from EPSRC grant: EP/T025964/1. A.~Gibberd, and L.~Mosley acknowledge funding from ESRC grant: ES/V006339/1. L.~Mosley acknowledges support from the STOR-i Centre for Doctoral Training and the Office for National Statistics.

\bibliographystyle{plainnat} 
\bibliography{refs}

\end{document}